\documentclass[final,3p,times]{elsarticle}

\usepackage{amssymb}

\usepackage{amsmath}

\newcommand{\bs}[1]{\boldsymbol{#1}}

\journal{Physica A}

\begin{document}

\begin{frontmatter}



\newcommand{\abs}[1]{\left\vert #1\right\vert}

\title{Consensus as cooling: a granular gas model for continuous opinions on structured networks}


\author{Carlos Uriarte}
\address{Área de Electromagnetismo, Universidad Rey Juan Carlos, 28933 Móstoles, Spain}
\ead{carlos.ugonzalez@urjc.es}
\author{Pablo Rodriguez-Lopez} 
\address{Área de Electromagnetismo \& Grupo Interdisciplinar de Sistemas Complejos, Universidad Rey Juan Carlos, 28933 Móstoles, Spain}
\ead{pablo.ropez@urjc.es}
\author{Nagi Khalil}
\address{Departamento de Física y Matemáticas \& Grupo Interdisciplinar de Sistemas Complejos, Universidad de Alcalá, 28805 Alcalá de Henares, Spain}
\ead{nagi.khalil@uah.es}

\begin{abstract}
  A continuous-opinion model accounting for the social compromise propensity is theoretically and numerically analysed. An agent's opinion is represented by a real number that can be changed through social interactions with her neighbours. The proposed dynamics depends on two fundamental parameters, \(\alpha\in[-1,1]\) and \(\beta\ge 0\). If an interaction takes place between two agents, their relative opinions decreases an amount given by \(\alpha\). The probability of two neighbours to interact is proportional to the \(\beta\)-power of their relative opinions. We unveil the behaviour of the system for all physical relevant values of the parameters and several representative interaction networks. When \(\alpha\in(-1,1)\) and \(\beta\ge 0\), the system always reaches consensus, with all agents having the mean initial opinion, provided the interaction network is connected. The approach to consensus can be characterized by means of the mean opinion and the temperature (or opinion dispersion) of each agent. Three scenarios have been identified. When the agents are well mixed, as with all-to-all interactions, a pre-consensus regime is seen, with all agents having zero mean opinion and the same temperature, following the Haff's law of granular gases. A similar regime is observed with Erdös-Rényi and Barabási-Albert networks: mean opinions are zero but agents with different degrees have different temperatures, though still following the Haff's law. Finally, the case of a square 2D lattice has been carefully analyzed, by starting from the derivation of closed set of hydrodynamic-like equations using the Chapman-Enskog method. For \(\alpha\) larger than a critical value, that depends on the system size, the system keeps spatially homogeneous, with zero mean opinions and equal temperatures, as they approach consensus. Below this critical line, the system explores states with spatially non-homogeneous configurations that evolve in time. Numerically, it is found that the main role of \(\beta\) is to change the local structure of the spacial opinion dispersion: while for \(\beta\) small enough the system keeps locally isotropic, as \(\beta\) increases, neighbouring agents with similar opinions tend to form local lineal structures. 
\end{abstract}




\end{frontmatter}



\section{Introduction \label{intro}}

An appropriate way to model social systems is to adopt a statistical physics perspective, assuming that the collective behaviour of individuals can be inferred from microscopic laws involving one or a few of them \cite{castellano2009statistical,jkedrzejewski2019statistical}. However, since individual behaviours may differ from one social context to another, models often assume different microscopic mechanisms. Two relevant examples are the Voter Model (VM) \cite{clifford1973model,holley1975ergodic} and the Deffuant Model (DM) \cite{deffuant2000mixing}, devoted to the understanding of the dynamics of opinion formation in social systems. Alongside, the Hegselmann-Krause model \cite{hegselmann2002opinion} represents another key paradigm for continuous opinions with bounded confidence.

In the VM, each individual can only adopt two possible states, while in the DM, the state is represented by a single continuous variable. Moreover, while in the VM the dynamics is driven by an asymmetric imitation process, where only one individual changes her opinion as a consequence of a dyadic social interaction, the DM describes social interactions as collision of two particles, with certain conservation quantities.

An important aspect to be taken into account upon proposing a model from a statistical physics perspective is that the nature of the social phenomena is inherently associated with non-equilibrium behaviour \cite{marro2005nonequilibrium,tauber2014critical}. Therefore, while models can describe social traits qualitatively, the concrete results may strongly depend on the details. That is to say, both the transient and steady-state properties of the social systems may be very sensitive not only to the possible states and the dynamics of change between them but also to the geometry of the social interactions, the so-called topology of the interaction network, among other factors. This has been exhaustively shown for the VM where the approach to a consensus state with all agents sharing the same opinion, for instance, depends critically on the dimensionality of the interaction network \cite{suchecki2005voter}, on the presence or absence of some free will upon changing opinion \cite{kirman1993ants}, on the existence of agents with specific features such as zealots and contrarians \cite{khalil2018zealots,khalil2019noisy,khalil2021zealots}, among others. Similar studies have been carried out within the DM and other continuous models, revealing how consensus, polarization, and fragmentation depend on the underlying network structure \cite{fortunato2004universality, lorenz2007power, fortunato2005vector, lorenz2007continuous}.

In \cite{khalil2021approach}, a new model for continuous opinion dynamics inspired by the phenomenology of granular gases \cite{brilliantov2010kinetic} was proposed and analyzed at the mean-field level where all agents interact with all others. Social interactions were described as inelastic collisions: once opinion and velocity are identified, the compromise propensity, or tendency to reduce the opinion distance, of the social interactions was modeled as a collision of two particles conserving linear momentum but reducing their relative velocities, hence dissipating kinetic energy. Moreover, the collision frequency was taken as the power \(\beta\ge 0\) of the difference of opinions, in a way different from a threshold model, including the DM as a special case. As a result, the model in \cite{khalil2021approach} always shows an evolution toward global consensus, as the DM with sufficiently high bound of confidence. More precisely, the so-called opinion temperature, a measure of the width of the distribution of opinions, was shown to be a monotonously decreasing function of time, following the so-called Haff's law of granular fluids \cite{haff1983grain}. However, the approach to consensus can be very different in both models. Namely, for the model in \cite{khalil2021approach} and for a given value of the dissipation and a sufficiently high exponent \(\beta\), two well-defined groups of agents with different mean opinions are formed before reaching consensus. This is shown as the distribution of opinions having two maxima which becomes closer as time rises. This novel behaviour is absent in the DM for a large bound of confidence, where the approach to global consensus is through a unimodal opinion distribution. The importance of the shape of the distribution of opinion beyond the steady state has been recognized to be also relevant in other voter-like models, see for instance \cite{kitching2025breaking}.

The main objective of this work is to unveil what happens with the model in \cite{khalil2021approach} beyond mean field, by considering non-trivial structures in the topology of the social interactions. Taking an appropriate limit, the DM for large enough bound of confidence will also be covered \cite{fortunato2004universality}. For regular networks, specifically for a lattice, our approach is, at some point similar to the approach in \cite{manacorda2016lattice,baldassarri2018hydrodynamics} for granular gases. Namely, starting from a stochastic description of the system, a macroscopic hydrodynamic-like description in terms of few quantities is derived at some limit. However, there are different ways to approach the hydrodynamics, with generally different results. Here we derive a macroscopic description using the Chapman-Enskog expansion adapted to granular gases \cite{brey1998hydrodynamics,khalil2014hydrodynamic,khalil2020unified}. 

The rest of the work is organized as follows. In Sec.~\ref{sec:model}, the continuous opinion model to be studied is introduced. A general description of the dynamics towards consensus, in terms of view macroscopic-like magnitudes, is given in Sec.~\ref{sec:approach}. The theoretical description is seen to be compatible to two different approaches to consensus: a homogeneous one where agents have the same statistical properties (same macroscopic properties and probability density), and also a non-homogeneous one with agents having some different properties. Section \ref{sec:lattice} analyses a set of agents interacting on a regular lattice. Starting from a simplified mesoscopic kinetic-like description, a closed hydrodynamic description is derived using the Chapman-Enskog method. The details of the derivation is in \ref{appen:1}. Spatially homogeneous solutions to the resulting description is identified and the study of their linear stability addressed. The theoretical results are compared and complemented with Monte Carlo numerical simulations in Sec.~\ref{sec:numsim}. The final Sec.~\ref{sec:diss} includes a discussion of the results and some concluding remarks. 

\section{Model\label{sec:model}}

The social context we model has a fixed number of agents, each one having an opinion on a given topic, represented by a real number. As usual, agents can change their opinions through social interactions that are assumed to take place through binary encounters. Moreover, social interactions are restricted to neighbours forming a social interaction network. The latter can have many different topological properties but is assumed to be connected, undirected, and static in time. Finally, the social interactions are assumed to be symmetric and to reflect a compromise propensity: both interacting agents change their opinions after the discussion with the same inertia and diminishing their differences. 

Mathematically, the proposed model is specified by: (a) fixing the number of agents \(N\), (b) the state of the system at a given time \(t\in\mathbb R\) with the vector \(\bs S_t=(s_{1},\dots,s_{N})\), where \(s_i\in\mathbb R\) is the opinion of the agent \(i\), and (c) the dynamics or time evolution of \(\bs S_t\). The dynamics depends on the structure of the social interactions which is given by a symmetric and undirected network, without loops or multilinks. A given node is always occupied by the same agent (there are no empty nodes). Using \(\Sigma=\{1,\dots,N\}\) to denote the set of nodes/agents, the social interactions of \(i\in \Sigma\) are restricted to her neighbouring nodes/agents, i.e. to the subset \(\mathcal V_i:=\{j\in\Sigma \,|\,j\text{ is a neighbour of }i\}\subset\Sigma\).

The state of the system \(\bs S_t\) can be seen as a stochastic process generated via the following Monte Carlo steps:
\begin{enumerate}
\item The initial state \(\bs S_0\) and the network are constructed. The former is generated by assigning opinions with Gaussian distributions, of zero means and unit variances, to each of the agents independently. Defining the network of interactions requires  the construction of the set of neighbours \(\{\mathcal V_{i}\}_{i\in\Sigma}\) using a specific network model. Here we consider the following ones: all-to-all, Erdös-Rényi, Barabási-Albert, and square 2D lattice.
\item For a given state \(\bs S_t\), a couple of agents \(i\) and \(j\), with opinions \(s_i\) and \(s_j\), are randomly selected with uniform probability. If they are not neighbours, no change is made. Otherwise, the agents \(i\) and \(j\) interact with probability
  \begin{eqnarray}
    \label{eq:freqsym}
    \min\left(1,\frac{|s_i-s_j|^\beta}{\Delta s_t^\beta}\right).
  \end{eqnarray}
  where \(\Delta s_t>0\) is chosen following different protocols, as specified below. If the interaction takes place, the agents change their opinions to the primed values as
  \begin{eqnarray}
    \label{eq:sip}
    && s_i'=b_{ij}s_i=s_i+\frac{1+\alpha}{2}(s_j-s_i), \\
    \label{eq:sjp}
    && s_j'=b_{ij}s_j=s_j-\frac{1+\alpha}{2}(s_j-s_i), 
  \end{eqnarray}
  where the last equalities define the collision operator \(b_{ij}\) as acting on the opinions \(s_i\) and \(s_j\).   
\item The time and the state are updated: \(t\to t+t_0/N\) and \(\bs S_t\to \bs S_{t+t_0/N}\), where \(t_0\) fixes the unit of time. 
\item If the new time is larger or equal to a final desired one, the simulations ends. Otherwise step 2 is run with the new state \(\bs S_{t+t_0/N}\).
\end{enumerate}

The dynamics depends on two important parameters. The parameter \(\alpha\in [-1,1]\) tunes the compromise propensity or tendency of the opinions to become more similar after the interaction: the interaction conserves the ``total opinion'' \(s_i'+s_j'=s_i'+s_j'\) but the opinion difference is reduced for \(\alpha\in(-1,1)\) as \(|s_i'-s_j'|=|\alpha||s_i-s_j|\). The case \(\alpha=0\) will not be considered in this work (see below). The parameter \(\beta\ge 0\) accounts for the influence of the opinion difference in the interaction probability.

The factor \(\Delta s_t\) can be chosen in different ways:
\begin{itemize}
    \item[-] For \(\Delta s_t\) taken time independent, typically \(\Delta s_t=2\sqrt{2}\), and  when \(|\alpha|\ne 1\) and \(\beta>0\), the quantity \({|s_i-s_j|^\beta}/{\Delta s_t^\beta}\) in Eq.~\eqref{eq:freqsym} becomes smaller and smaller as time rises. 
    \item[-] An alternative approach uses an estimation of the typical opinion differences at a given time. This can be done by running the Monte Carlo algorithm with a constant \(\Delta s_t\) once and taking the maximum opinion difference found at each time step to define a new \(\Delta s_t\) for future simulations. The resulting \(\Delta s_t\) is time dependent so that the probability of accepting a social interaction remains (almost) constant.
\end{itemize}

\subsection{Distribution of opinions}

The previous model can be described theoretically starting from a master equation. Following the procedure of \cite{khalil2021approach}, it is not difficult to see that, in the continuum-time limit, the distribution of opinions of the agent \(i\) at time \(t\), \(p_i(s_i,t)\), or the probability density of finding the agent \(i\) with opinion \(s_i\) at time \(t\), obeys
\begin{equation}
\label{eq:pi}
  \partial_tp_i(s_1,t)=\frac{1}{Nt_0\Delta s_t^\beta}\int ds_2\, (|\alpha|^{-1}b_{12}^{-1}-1)|s_1-s_2|^\beta \sum_{j\in\mathcal V_i} p_{ij}(s_1,s_2,t),
\end{equation}
where it is assumed that the integral runs over all \(s_2\in\mathbb R\), \(p_{ij}(s_1,s_2,t)\) is the joint probability density of the agents \(i\) and \(j\) to have opinions \(s_1\) and \(s_2\), respectively, and \(b_{12}^{-1}\) is the inverse of \(b_{12}\), acting on any function \(f(s_1,s_2)\) as \(b_{12}^{-1}f(s_1,s_2)=f(b_{12}^{-1}s_1,b_{12}^{-1}s_2)\) with 
\begin{eqnarray}
    && b_{12}^{-1} s_1=s_1+\frac{1+\alpha}{2\alpha}(s_2-s_1), \\
    && b_{12}^{-1} s_2=s_2-\frac{1+\alpha}{2\alpha}(s_2-s_1). 
\end{eqnarray}
Since the operator \(b_{12}^{-1}\) is defined only when \(\alpha\ne 0\), the case \(\alpha=0\) is not addressed in this work; the interested reader is referred to Ref.~\cite{carro2013role}.

Equation \eqref{eq:pi} cannot be solved to obtain the opinion distribution \(p_i(s_i,t)\) without knowing the correlations between neighbouring agents, \(p_{ij}(s_1,s_2,t)\), unless we consider additional equations or use some approximations. However, it constitutes our starting point for the forthcoming analysis. Approximations are considered later. 

\subsection{Mean opinion and temperature}

Beyond the distribution of opinions, it is useful to introduce new scalar quantities that give information about the social behaviour of agents, such as the mean opinion \(S_i\) and the opinion temperature \(T_i\) of a given agent \(i\). They are defined through the first two moments of the distribution of opinions: 
\begin{eqnarray}
    && S_i(t)=\int ds\, s p_i(s,t), \\
    && T_i(t)=\int ds\, (s-S_i)^2 p_i(s,t)=\int ds\, s^2 p_i(s,t)-S_i^2(t).
\end{eqnarray}
Multiplying Eq.~\eqref{eq:pi} by \(s_1\) and \((s_1-S_i)^2\), integrating over \(s_1\), and after some computations, see Ref.~\cite{khalil2021approach}, we obtain
\begin{eqnarray}
\label{eq:Sit}
  && \frac{d}{dt}S_i(t)=\frac{1+\alpha}{2Nt_0\Delta s_t^\beta}\iint ds_1 ds_2|s_1-s_2|^\beta(s_2-s_1) \sum_{j\in\mathcal V_i}p_{ij}(s_1,s_2,t), \\
  \label{eq:Tit}
  && \frac{d}{dt}T_i(t)+2S_i(t)\frac{d}{dt}S_i(t)=\frac{1+\alpha}{Nt_0\Delta s_t^\beta}\iint ds_1 ds_2|s_1-s_2|^\beta(s_2-s_1)\left[s_1+\frac{1+\alpha}{4}(s_2-s_1)\right] \sum_{j\in\mathcal V_i} p_{ij}(s_1,s_2,t).
\end{eqnarray}

Global mean opinion and temperature provide information about the system as a whole. The former can be defined as the average value of \(S_i\) over all the agents:
\begin{equation}
    S(t)=\frac{1}{N}\sum_{i=1}^N S_i.
\end{equation}
The definition of the global temperature is not unique. Instead of defining it as \(\frac{1}{N}\sum\limits_{i=1}^N T_i\), we take
\begin{equation}
    T(t)=\frac{1}{N}\sum_{i=1}^N\int ds\, (s-S)^2p_i(s,t),
\end{equation}
which is defined in terms of the global mean opinion \(S\). By means of Eq.~\eqref{eq:pi}, following \cite{khalil2021approach}, the equations for \(S\) and \(T\) read 
\begin{eqnarray}
\label{eq:globals}
    && \frac{d}{dt}S(t)=0, \\
    \label{eq:globalT}
    && \frac{d}{dt}T(t)=-\zeta(t) T(t),
\end{eqnarray}
where 
\begin{equation}
\label{eq:coolingr}
    \zeta(t)=\frac{1-\alpha^2}{4N^2t_0\Delta s_t^\beta T(t)}\sum_{i=1}^N\sum_{j\in\mathcal V_i}\iint ds_1 ds_2|s_1-s_2|^{\beta+2} p_{ij}(s_1,s_2,t)
\end{equation}
is the so-called cooling rate. It measures how fast is the global approach to consensus at a given time. 

In summary, the proposed opinion model assumes each agent can have an opinion represented by a real number that can change through stochastic binary social interactions between neighbours. The evolution of the system depends importantly on two parameters: while \(\alpha\) tunes the effects of the interactions, \(\beta\) modulates their frequency. Beyond the microscopic evolution of the system, a mesoscopic and macroscopic ones are also to be used. The latter takes the probability density \(p_i(s,t)\) of the agent \(i\) to have an opinion \(s\) at time \(t\) as the fundamental quantity, while the latter only considers the first two moments of \(p_i\): the mean opinion \(S_i(t)\) and temperature \(T_i(t)\), or even their global versions. In the subsequent sections, we provide valuable information of the system by working on the new descriptions. 

\section{Approach to consensus: scaling and Haff's law \label{sec:approach}}

A first important property of the proposed model is that it always describes an approach to consensus, provided the interaction network is connected and \(\alpha\in(0,1)\), see Ref.~\cite{khalil2021approach}. Beyond this result, the main objective of the present work is to describe how is the system evolution under different relevant social contexts.

That the system ends up in global consensus can be seen using Eqs.~\eqref{eq:globals} and \eqref{eq:globalT}. From Eq.~\eqref{eq:globals} it follows that the global opinion is conserved. This is the macroscopic manifestation of the microscopic conservation of opinions of the collision rule \eqref{eq:sip}-\eqref{eq:sjp}. Taking into account that initially \(S_i=0\) by construction, we have 
\begin{equation}
    S(t)=0.
\end{equation}
However, the previous relation does not imply that \(S_i(t)=0\) for all \(t\), as seen in Sec.~\ref{sec:lattice}. 

When \(|\alpha|\ne 1\) and for some initial opinion dispersion, \(T>0\), the cooling rate Eq.~\eqref{eq:coolingr} is a positive function of time, \(\zeta(t)>0\). This implies, from Eq.~\eqref{eq:globalT}, that the global temperature \(T\) is a decreasing function of time. This is a consequence of the microscopic collision rule \eqref{eq:sip}-\eqref{eq:sjp} being dissipative. Moreover, the only steady microscopic state corresponds to \(T=0\), i.e. no opinion dispersion \(s_i=S=0\):
\begin{equation}
  \label{eq:consensus}
    T(t)\to 0\quad \Rightarrow\quad s_i\to S=0.
\end{equation}
Any other state with \(T>0\) have \(\zeta>0\) and has a decreasing \(T\).

Hence, for \(|\alpha|\ne 1\) the dynamics always drives the system towards the global consensus with \(S=0\). With a similar reasoning, we can also infer that, if initially \(S(0)\ne 0\), the system also reaches consensus with \(s_i=S(0)\). For \(|\alpha|=1\) the dynamics conserves the mean opinion and the temperature: for \(\alpha=-1\) there is no evolution at all (interactions do not change the opinions and all agents keep their initial opinions) while for \(\alpha=1\) interactions only interchange the opinions of the two interacting agents. 

Next, we focus on the way the system approaches consensus. This is done by identifying different social conditions: homogeneous and non-homogeneous (including weak and strong). 

\subsection{Homogeneous approach to consensus}

The simplest scenario corresponds to a homogeneous evolution toward consensus. Here we mean that the statistical property of any given agent is the same as any other, at any time. More precisely, we consider a (mesoscopic) state to be homogeneous when
\begin{equation}
  \label{eq:homcond}
    p_{ij}(s_1,s_2,t)=p_{ji}(s_1,s_2,t), \quad i\in\Sigma,\,j\in\mathcal V_i. 
\end{equation}
If we now integrate the previous relation with respect to \(s_2\) we have
\begin{equation}
  p_i(s_1,t)=p_j(s_1,t), \quad i\in\Sigma,\,j\in\mathcal V_i.
\end{equation}
Taking into account that we are considering a connected network of neighbours, the previous relation implies that all agents share the same distribution of opinions:  
\begin{equation}
  \label{eq:pis1ps1}
  p_i(s_1,t)=p(s_1,t), \quad i\in\Sigma.
\end{equation}

It is not easy to verify when condition \eqref{eq:homcond} is verified. In general, it depends on the full dynamics: the interaction structure plus the values of the parameters \(N,\,\alpha\), and \(\beta\). However, for a well-mixed population, as is the case with all-to-all interactions, the system is expected to reach homogeneous states for all the parameters of the system (except maybe for \(\alpha=-1\)). This is because, under well-mixed conditions, all agents see (almost) the same neighbourhood and interchanging them makes no important difference to the state nor to the dynamics, at least for \(N\) large enough. Hence, homogeneous condition \eqref{eq:homcond} seems natural in this case. 

The homogeneous condition \eqref{eq:homcond} strongly restricts the possible macroscopic behaviour of the system. Using Eq.~\eqref{eq:homcond} with Eqs.~\eqref{eq:Sit}-\eqref{eq:Tit}, it is readily seen that \(S_i\) and \(T_i\) obeys the same equations as \(S\) and \(T\), respectively. Additionally, since global and local quantities coincide at \(t=0\), we infer that
\begin{equation}
  \label{eq:sis0}
  S_i=S=0, \quad i\in\Sigma
\end{equation}
and
\begin{equation}
  \label{eq:titt}
  T_i(t)=T(t),\quad i\in\Sigma
\end{equation}
for all \(t\ge 0\). That is, the system stays homogeneous within the macroscopic description. 

The relations \eqref{eq:sis0} and \eqref{eq:titt} do not preclude any form of the opinion distribution \(p(s,t)\), beyond \(\int ds\, p(s,t)=0\) imposed by \(S_i=0\). As a matter of fact, \(p(s,t)\) can have many different shapes, including unimodal and bimodal, as shown in Ref.~\cite{khalil2021approach} in the all-to-all case. In the unimodal case the opinions distribute around a maximum of \(p(s,t)\) located at the zero opinion, while with bimodal \(p(s,t)\) two sub-populations with opposite mean opinions can be identified. However, bimodality is not in contradiction with homogeneity, since the two identified sub-populations are, on the one hand, uniformly distributed throughout the system and, on the other hand, all agents, at a given time, belong to one of the two sub-populations with the same probability.

\subsection{Scaling property}

Under the (macroscopic) homogeneous conditions \eqref{eq:sis0} and \eqref{eq:titt}, and by analogy with other particle systems \cite{khalil2021approach,khalil2018generalized}, it is natural to identify the global temperature \(T\) as the only relevant magnitude for measuring the opinions. With the previous assumption, and by means of a dimensional analysis, we find the following scaling property:
\begin{equation}
\label{eq:scalpij}
    p_{ij}(s_1,s_2,t)=[2T(t)]^{-1} \phi_{ij}(c_1,c_2),\quad i,j\in\Sigma,
\end{equation}
where \(\phi_{ij}\) is a scaling function and 
\begin{equation}
    c=\frac{s}{\sqrt{2T}}
\end{equation}
is a new opinion variable.

Integrating Eq.~\eqref{eq:scalpij} over \(s_2\), a similar scaling property is found for the opinion distribution:
\begin{equation}
  \label{eq:scalingphi}
    p_i(s,t)=[2T(t)]^{-\frac12}\phi_i(c),\quad i\in \Sigma
\end{equation}
where 
\begin{equation}
  \label{eq:phii}
    \phi_i(c)=\int dc_2\, \phi_{ij}(c,c_2), \qquad i,j\in\Sigma.
\end{equation}
Moreover, using Eq.~\eqref{eq:pis1ps1}, we have equality for all scaling functions \(\phi_i(c)\) in \eqref{eq:scalingphi}:
\begin{equation}
  \phi_i(c)=\phi(c),\quad i\in\Sigma.
\end{equation}
This property can also be seen from \eqref{eq:phii} noting that Eq.~\eqref{eq:homcond} and the scaling property \eqref{eq:scalpij} imply \(\phi_{ij}(c_1,c_2)=\phi_{ji}(c_1,c_2)\). A similar property does not hold for \(\phi_{ij}\) in general. Namely, \(\phi_{ij}(c_1,c_2)\ne \phi_{ik}(c_1,c_2)\) for \(j\ne k\), in general, even though \(\phi_j=\phi_k\). The same is also true in homogeneous liquid system \cite{santos2016concise}. 

Using the hypothesis \eqref{eq:scalpij} with the cooling rate Eq.~\eqref{eq:coolingr} we obtain a more explicit equation for the global temperature, the so-called Haff's law:
\begin{equation}
  \label{eq:haffl}
  \frac{d}{dt}T(t)=-\frac{\tilde \zeta}{t_0} \left(\frac{2\sqrt{2T(t)}}{\Delta s_t}\right)^\beta T(t),
\end{equation}
with
\begin{equation}
  \label{eq:tzeta}
  \tilde\zeta =\frac{1-\alpha^2}{2^{\beta+1}N^2}\iint dc_1 dc_2|c_1-c_2|^{\beta+2} \sum_{i=1}^N\sum_{j\in\mathcal V_i}\phi_{ij}(c_1,c_2)\ge 0
\end{equation}
a time independent coefficient. The Haff's law implies two important features of the time dependence of the global temperature. On the one hand, if \(\Delta s_t\) is a constant, the time evolution of \(T\) is given by 
\begin{eqnarray}
  \label{eq:ttbet0}
  && T(t)=T_0e^{-\frac{\tilde\zeta}{t_0} t},\quad \beta=0,\\
  \label{eq:ttbet>0}
  && T(t)=\left(T_0^{-\frac{\beta}{2}}+\frac{\beta \tilde \zeta}{2t_0}t\right)^{-\frac{2}{\beta}}, \quad \beta>0,
\end{eqnarray}
with \(T_0\) the initial temperature. Interestingly, \(T(t)\propto t^{-\frac{2}{\beta}}\) for \(t\gg t_0\) when \(\beta>0\), with the exponent being independent of \(\alpha\). On the other hand, if we make the following change of the time variable
\begin{eqnarray}
  \label{eq:tiempotau}
  d\tau=\frac{T_h^{\frac{\beta}{2}}(t)}{\Delta s_t^\beta t_0}dt
\end{eqnarray}
even if \(\Delta s_t\) is time dependent, then 
\begin{equation}
  \label{eq:ttexp}
  T(\tau)= T_0\exp\left(-\frac{\tilde\zeta}{2^{\frac{\beta+2}{2}}} \tau\right)
\end{equation}
for all \(\beta\ge 0\). The coefficient of the exponential decay of \(T\) is directly given by \(\tilde \zeta\), a function of both \(\alpha\) and \(\beta\). 

An important fact that supports the relevance of the scaling relation \eqref{eq:scalpij} is its compatibility with Eq.~\eqref{eq:pi} for the opinion distribution \(p_i\). That is, the (mesoscopic) dynamics is compatibly with the proposed scaling. Namely, when Eqs.~\eqref{eq:scalpij} and \eqref{eq:scalingphi} are used with Eq.~\eqref{eq:pi}, and the Haff's law \eqref{eq:haffl} are taken into account, a consistent equation for \(\phi\), with no explicit time dependence, is obtained:
\begin{equation}
  \label{eq:kineticeqesc}
  2^{\beta-1}N\tilde \zeta\partial_{c_1}[c_1\phi(c_1)]=\int dc_1\, (|\alpha|^{-2}b_{12}^{-1}-1)|c_1-c_2|^\beta\sum_{j\in\mathcal V_i}\phi_{ij}(c_1,c_2),
\end{equation}
where \(\tilde \zeta\) is a functional of \(\phi_{ij}\) and \(b_{12}^{-1}\) acts on \(c_1\) and \(c_2\) as with \(s_1\) and \(s_2\), respectively.

Equation \eqref{eq:kineticeqesc} was approximately solved in Ref.~\cite{khalil2021approach} taking a mean-field-like limit, valid for all-to-all interaction and for large populations. Under these conditions, it was numerically shown that the system always reaches a final state that fits the (macroscopic) homogeneous conditions \eqref{eq:sis0} and \eqref{eq:titt} with a velocity distribution having the scaling property \eqref{eq:scalingphi}. Moreover, for a given \(N\), the \((\alpha,\beta)\) space of parameters splits into two regions characterized by the scaling function \(\phi\) being unimodal or bimodal, with a global symmetry between \(\alpha>0\) and \(\alpha<0\). The analysis made so far in this subsection makes it possible to have a similar results beyond the all-to-all scenario, with eventual important contributions from the structure of the interactions. 

In summary, the scaling property \eqref{eq:scalpij} makes all the time dependence of the distributions \(p_{ij}\) and \(p_i\) to take place only through their dependence on the global temperature \(T\), in such a way that if opinions are measured in units of \(\sqrt{2T}\), then the scaled distributions \(\phi_{ij}\) and \(\phi_i\) become time independent. The scaling is expected to occur only under restricted dynamical conditions (topology of the interactions and parameters of the system) to be investigated, and only after a transient time, long enough so that the system can forget the initial conditions. Homogeneity and scaling manifest macroscopically through relations \eqref{eq:sis0} for the mean opinions and \eqref{eq:titt} for the temperatures, together with the Haff's law \eqref{eq:haffl} for the global temperature  that gives rise to explicit time behaviours \eqref{eq:ttbet0}, \eqref{eq:ttbet>0}, and \eqref{eq:ttexp}. All the previous points are strong theoretical assumptions/predictions to be verified numerically, see Sec.~\ref{sec:numsim}.

\subsection{Weak non-homogeneous approach to consensus}

The homogeneity condition is very restrictive and may not be verified beyond the all-to-all scenario (see however the analysis of Sec.~\ref{sec:lattice}). Nevertheless, even through the structure of the interactions prevents condition \eqref{eq:scalpij} to be fulfilled, i.e. under non-homogeneous conditions, the system may reach, after a transient, a state where the global temperature still serves as the fundamental gauge. This is expected to occur at least for interaction structures ``similar'' to the all-to-all one. It has also been observed in granular systems as well, see Refs.~\cite{brey2010critical,brey2011equilibration}. 

More specifically, under non-homogeneous, but ``close'' to homogeneous conditions (we refer to them as weak non-homogeneous) we make the following hypothesis:
\begin{itemize}
\item[-] the mean opinion of agents tend in the long time to zero,
  \begin{equation}
    \label{eq:nohom1}
    S_i(t)=S(t)=0,\quad i\in\Sigma,
  \end{equation}
  as for homogeneous conditions, but
\item[-] the local and global temperatures differ, in general, with
  \begin{equation}
    \label{eq:nohom2}
    \frac{d}{dt}\left(\frac{T_i}{T}\right)=0,\quad i\in\Sigma.
  \end{equation}
  That is, \(T_i/T\) can change from one agent to another, but the time dependence of the local opinions is given by the global one.
\item[-] Moreover, the scaling properties of the distributions \eqref{eq:scalpij} and \eqref{eq:scalingphi} also hold.
\end{itemize}

With the new hypothesis, the Haff's law \eqref{eq:haffl} is still valid. Interestingly, the structure of the interactions only affects the value of the scaling cooling rate in \eqref{eq:tzeta}. Again, we have important theoretical predictions that have to be compared against numerical simulations, see Sec.~\ref{sec:numsim}.

In the next section we analyse the case of a system with social interactions forming a lattice, an instance of a spatial network. This is an appropriate case where the scaling hypothesis based on the homogeneity conditions may fail. 

\section{Social interactions on a lattice: Spatial stability \label{sec:lattice}}

The case when the interaction network forms a hypercubic regular lattice in \(d\) dimensions,  with each agent having \(2d\) neighbours, is analyzed. Under these conditions, to each agent \(i\) can be given a fixed position \(\bs r_i\in\mathbb R^d\). It is then convenient to slightly change the notation by replacing the suffix indicating the agent by her respective position vector as:
\begin{eqnarray}
  \label{eq:iden1}
  &&p_i(s,t)\to p(\bs r_i,s,t), \\
  \label{eq:iden2}
  &&S_i(t) \to S(\bs r_i,t),\\
  \label{eq:iden3}
  &&T_i(t)\to T(\bs r_i,t).
\end{eqnarray}
Without loss of generality, we measure distances in units of the lattice spacing.

Additionally, we take the following mathematical statements as true:
\begin{enumerate}
\item The domain of \(p(\cdot,s,t)\) is extended to include all the space points in between the agents positions in such a way that the resulting function is regular enough. As a result, the local mean opinion \(S(\bs r,t)\) and temperature \(T(\bs r,t)\) are also extended. 
\item The distribution of opinions \(p(\bs r,s,t)\) is a smooth function of the position such that
  \begin{equation}
    p(\bs r_j,s,t)\simeq p(\bs r_i,s,t)\pm \hat e_k\cdot \partial_{\bs r}p(\bs r_i,s,t)+\frac{1}{2}(\hat e_k\cdot \partial_{\bs r})^2p(\bs r_i,s,t).
  \end{equation}
  when \(i\) and \(j\) are neighbours and \(\bs r_j-\bs r_i=\pm \hat e_k\) where \(\hat e_k\) is the unit vector along the segment joining the two agents. 
\item The statistical correlations among interacting neighbours are removed (molecular chaos): 
  \begin{equation}
    p_{ij}(s_1,s_2,t)\simeq p(\bs r_i,s_1,t)p(\bs r_j,s_2,t).
  \end{equation}
\end{enumerate}

A first important consequence of the previous assumptions is a simplification of the right-hand side of Eq.~\eqref{eq:pi}:
\begin{equation}
  \label{eq:approx1}
  \sum_{j\in\mathcal V_i} p_{ij}(s_1,s_2,t)\simeq p(\bs r_i,s_1,t) \sum_{j\in\mathcal V_i} p(\bs r_j,s_2,t)\simeq p(\bs r_i,s_1,t)[2dp(\bs r_i,s_2,t)+\nabla^2 p(\bs r_i,s_2,t)].
\end{equation}
Note that all previous results hold because the interaction rate, proportional to \(|s_1-s_2|^\beta\), does not depend on the agent space position, but exclusively on the opinion difference. This is an important difference with respect to a collision model of, say, hard grains, as discussed in \cite{baldassarri2018hydrodynamics}. Another important difference with physical (granular) systems has to be with the dimensionality: while space has dimension \(d\ge 1\), the opinion of an agent is always one dimensional. 

With the approximation \eqref{eq:approx1}, the exact Eq.~\eqref{eq:pi} for the distribution of opinions becomes a closed, though approximate equation for \(p(\bs r,s,t)\). It can be written as 
\begin{eqnarray}
  \label{eq:kineticeq}
  \partial_{t}p(\bs r_i,s_1,t)&\simeq& \int ds_2\, (|\alpha|^{-1}b_{12}^{-1}-1)\pi(s_1,s_2)p(\bs r_i,s_1,t)\left[2dp(\bs r_i,s_2,t) +\nabla^2p(\bs r_i,s_2,t)\right] 
\end{eqnarray}
where the time variable \(t\) has been redefined to absorb the factor \(Nt_0\), \(t/(Nt_0)\to t\), and the rate function \(\pi(s_1,s_2)\) is
\begin{equation}
  \label{eq:pis1s2}
  \pi(s_1,s_2)=\frac{|s_1-s_2|^\beta}{\Delta s_t^\beta}.
\end{equation}
Equation \eqref{eq:kineticeq} is a kinetic equation to be solved with an initial condition and the normalization condition
\begin{equation}
  \int ds\, p(\bs r,s,t)=1.
\end{equation}
Moreover, the kinetic equation \eqref{eq:kineticeq} is restricted to small gradients, in such a way that
\begin{equation}
  \label{eq:condconst}
  2dp(\bs r_i,s_2,t) +\nabla^2p(\bs r_i,s_2,t)\ge 0
\end{equation}
for all \(\bs r_i\), \(s_2\), and \(t\).

An alternative way of writing the kinetic equation~\eqref{eq:kineticeq}, which better reflects its symmetries, is
\begin{eqnarray}
  \label{eq:kineticeq2}
  \partial_{t}p(\bs r_i,s_1,t)&\simeq& \int ds_2\, (|\alpha|^{-1}b_{12}^{-1}-1)\pi(s_1,s_2)\left[2dp(\bs r_i,s_2,t)p(\bs r_i,s_1,t)-\nabla p(\bs r_i,s_1,t)\cdot \nabla p(\bs r_i,s_2,t)\right] \nonumber \\ &&+ \nabla\cdot \int ds_2\, (|\alpha|^{-1}b_{12}^{-1}-1)\pi(s_1,s_2)p(\bs r_i,s_1,t)\nabla p(\bs r_i,s_2,t).
\end{eqnarray}
It will be analyzed next, with most of the results being valid for generic rates \(\pi(s_1,s_2)\) that are even functions of the opinion difference \(s_1-s_2\).

\subsection{Hydrodynamics}

In order to analyse the spatial structure of the solutions, we focus on a macroscopic, hydrodynamic-like description, in terms of the first moments of the opinion distributions. In the present context, the relevant hydrodynamic fields are the local mean opinion \(S(\bs r,t)\) and the opinion temperature \(T(\bs r,t)\), defined in Eqs.~\eqref{eq:Sit} and \eqref{eq:Tit} and here written with the spatial notation. Using a standard methodology in Kinetic Theory, see the details in \ref{appen:11}, the equations for the two fields obtained from the kinetic equation \eqref{eq:kineticeq2} read
\begin{eqnarray}
  \label{eq:balanceS}
  && \partial_t S(\bs r,t)=\frac{1+\alpha}{2} \nabla \cdot \iint ds_1ds_2(s_2-s_1)\pi(s_1,s_2)p(\bs r,s_1,t)\nabla p(\bs r,s_2,t), \\ 
  \label{eq:balanceT}
  &&\partial_tT(\bs r,t)+2S(\bs r,t)\partial_tS(\bs r,t)\nonumber \\
  && \qquad  = -\frac{(1-\alpha^2)}{4}\iint ds_1ds_2(s_2-s_1)^2\pi(s_1,s_2)\left[2dp(\bs r,s_1,t) p(\bs r,s_2,t)-\nabla p(\bs r,s_1,t) \cdot \nabla p(\bs r,s_2,t)\right] \nonumber \\ && \qquad\quad  +(1+\alpha)\nabla\cdot \iint ds_1ds_2(s_2-s_1)\left[s_1+\frac{1+\alpha}{4}(s_2-s_1)\right]\pi(s_1,s_2) p(\bs r,s_1,t)\nabla  p(\bs r,s_2,t).
\end{eqnarray}
As is apparent, the set of equations for \(S\) and \(T\) is not closed since new unknown quantities appear. An exception is when \(\pi(s_1,s_2)\) is a constant function, independent of the opinions, as when \(\beta=0\). This case is analysed in \ref{appen:12}.

In a general situation, whether \(\pi\) depends on the opinions or not, we can use the Chapman-Enskog method to obtain closed hydrodynamic equations. This is a perturbative method that provides both, solutions to the kinetic equation in terms of the hydrodynamic equations and closed hydrodynamic equations at the desired order in the gradients. It has been long used with molecular and similar systems, see Ref.~\cite{chapman1990mathematical}, and also with granular gases, see Ref.~\cite{brey1998hydrodynamics}. Up to the second order in the gradients (and neglecting gradient contributions to the cooling rate), see the details in \ref{appen:13}, the method provides the following hydrodynamic description
\begin{eqnarray}
  \label{eq:diffS}
  \partial_t S(\bs r,t) &\simeq& \nabla \cdot [D(\bs r,t)\nabla S(\bs r,t)], \\
  \label{eq:eqtemp}
  \partial_tT(\bs r,t)& \simeq& -\left\{\zeta(\bs r,t)T(\bs r,t)-\eta_S(\bs r,t)\left[\nabla S(\bs r,t)\right]^2-\eta_T(\bs r,t)\left[\nabla T(\bs r,t)\right]^2\right\}+\nabla\cdot\left[\kappa(\bs r,t)\nabla T(\bs r,t)\right],
\end{eqnarray}
where the transport coefficients \(D\), \(\zeta\), \(\eta_S\), \(\eta_T\), and \(\kappa\) depend on \(\bs r\) and \(t\) through \(T\) and, eventually, also on \(t\) through \(\Delta s_t\). Approximate explicit expressions of them are given in Eqs.~\eqref{eq:zeta} and \eqref{eq:kappa}--\eqref{eq:etaT}. They coincide with the exact ones when \(\beta=0\).

Two important observations are in order. First, all transport coefficients are positive for all values of the parameters except \(\eta_S\) and \(\kappa\) that can be negative for \(|\alpha|\ne 1\) and \(\beta\) big enough. The case of \(\kappa\) is particularly surprising since it can be interpreted as a thermal conductivity (the absence of a streaming term in the kinetic equation may be the reason for this result). For \(\alpha=1\) or \(\beta=0\) all coefficients are positive. Second, Eqs.~\eqref{eq:diffS}--\eqref{eq:eqtemp} are consistent with the results of Sec.~\ref{sec:approach}: the global mean opinion is zero and the global mean temperature is a decreasing function of time. This is evident for Eq.~\eqref{eq:diffS}: after integrating over all \(\bs r\), using the periodic or reflecting boundary conditions (i.e. assuming the system is isolated), and using the initial condition \(S(0)=0\). Doing the same with Eq.~\eqref{eq:eqtemp}, we have
\begin{equation}
  \frac{d}{dt}T(t)=-\int d\bs{r}\, \left\{\zeta(\bs r,t)T(\bs r,t)-\eta_S(\bs r,t)\left[\nabla S(\bs r,t)\right]^2-\eta_T(\bs r,t)\left[\nabla T(\bs r,t)\right]^2\right\}\le 0,
\end{equation}
where the last inequality holds after taking into account that gradients must be small, or more precisely Eq.~\eqref{eq:condconst}. When \(\eta_S\ge 0\), recalling that \(\zeta,\eta_T\ge 0\), the previous expression tells us that any spatial dependence of \(S\) or \(T\) (nonzero gradients) effectively reduces the cooling rate, which makes the approach to consensus slower.

Having a closed set of hydrodynamic equations, Eqs.~\eqref{eq:diffS} and \eqref{eq:eqtemp}, we can proceed by analysing the existence of spatially homogeneous solutions that could correspond to the homogeneous states identified in Sec.~\ref{sec:approach}. Moreover, by studying the linear stability of the solutions, we can identify possible new mechanisms of approaching consensus.

\subsection{Spatially homogeneous approach to consensus }

The hydrodynamic equations admit spatially homogeneous solutions, \(S_h(t)\) and \(T_h(t)\). They correspond to the homogeneous approach to consensus previously identified under more general conditions in Sec.~\ref{sec:approach}. They can also be identified with the so-called homogeneous cooling state of a granular gas (HCS) \cite{khalil2018generalized}. 

By seeking solutions to the hydrodynamic equations \eqref{eq:diffS}--\eqref{eq:eqtemp} of the form \(S(\bs r,t)=S_h(t)\) and \(T(\bs r,t)=T_h(t)\), we have
\begin{eqnarray}
  && \partial_t S_h=0, \\
  && \partial_tT_h\simeq -\zeta(t)T_h,
\end{eqnarray}
with
\begin{eqnarray}
  \zeta(t)\simeq \frac{2^{\beta+1}d(1-\alpha^2)}{\sqrt{\pi}}\Gamma\left(\frac{\beta+3}{2}\right)\frac{[T_h(t)]^{\frac{\beta}{2}}}{\Delta s_t^\beta}.
\end{eqnarray}
Using the initial value \(S_h(0)=0\), we obtain \(S_h=0\), as expected. Moreover, the  granular temperature \(T_h(t)\) obeys the Haff's law \eqref{eq:haffl} with
\begin{equation}
  \tilde \zeta\simeq \frac{d(1-\alpha^2)}{2^{\frac{\beta-2}{2}}N\sqrt{\pi}}\Gamma\left(\frac{\beta+3}{2}\right).
\end{equation}

\subsection{Stability of the HCS}

We now seek solutions to the hydrodynamic equations of the form 
\begin{eqnarray}
  && S(\bs r,t)=S_h(t)+\sqrt{2T_h(t)}\sigma(\bs r,t), \\
  && T(\bs r,t)=T_h(t)[1+\theta(\bs r,t)],
\end{eqnarray}
where \(\sigma \) and \(\theta\) are assumed to depend on space and time. Substituting the new expressions into the Navier-Stokes equations \eqref{eq:diffS}--\eqref{eq:eqtemp}, keeping terms up to linear order in \(\sigma\) and \(\theta\), using the equation for \(T_h(t)\), and after some manipulations, we arrive at the equation for \(\sigma\):
\begin{equation}
  \label{eq:linsig}
  \partial_\tau \sigma(\bs r,\tau)\simeq  \frac{2^{\beta}d(1-\alpha^2)}{\sqrt{\pi}}\Gamma\left(\frac{\beta+3}{2}\right)\left[\sigma(\bs r,\tau) + k_\sigma^{-2}\nabla^2 \sigma(\bs r,\tau)\right],
\end{equation}
where \(\tau\) is the time variable given by Eq.~\eqref{eq:tiempotau} and 
\begin{equation}
  k^2_\sigma=d(1-\alpha)
\end{equation}
defines a wave vector \(k_\sigma\). As for the temperature, we have
\begin{equation}
  \label{eq:linthe}
  \partial_\tau \theta \simeq -\frac{2^{\beta}(1-\alpha^2)}{\sqrt{\pi}}\beta\Gamma\left(\frac{\beta+3}{2}\right)\left[\theta(\bs r,\tau)+k_\theta^{-2} \nabla^2 \theta(\bs r,\tau)\right],
\end{equation}
with a new wave vector \(k_\theta\) given by
\begin{equation}
  k_\theta^2=\frac{4d(1-\alpha)\beta}{2\beta-(1+\alpha)(2+\beta)}.
\end{equation}

Taking the Fourier transform of Eqs.~\eqref{eq:linsig} and \eqref{eq:linthe}, it is readily seen that the homogeneous solution is stable provided all possible wave vectors \(k\) fulfill the following conditions simultaneously:
\begin{eqnarray}
  \label{eq:condksig}
  &&k^2>k_\sigma^2, \\
  \label{eq:condkthe}
  &&k^2<k^2_\theta,\quad \text{or}\quad k_\theta^2<0.
\end{eqnarray}
For the first one to be verified, it is enough to have \(2\pi/L>k_\sigma\), where \(L\) is the ``smallest'' length of the system. If it is a hypercube in \(d\) dimensions, then \(N=L^d\) and the condition \eqref{eq:condksig} reads
\begin{equation}
  \label{eq:crital}
  \alpha>1-\frac{4\pi^2}{dN^{\frac{2}{d}}}.
\end{equation}
As for the Eq.~\eqref{eq:condkthe}, it is enough to impose \(k_{\max}<k_\theta\) or \(k^2_\theta<0\). In the first case, taking \(k_{\max}=2\pi\) as the maximum possible wave vector, we have \(\beta(1-8\pi d)(1-\alpha)>2(1+\alpha)\) which is impossible for \(\alpha\in[-1,1]\) and \(\beta\ge 0\). Hence, the condition \eqref{eq:condkthe} is equivalent to \(k_\theta^2<0\) or
\begin{equation}
  \label{eq:critbe}
  \beta< \frac{2(1+\alpha)}{1-\alpha}.
\end{equation}

With the previous results we can compose a picture of the evolution of the agents towards consensus. Fixing the number \(N\) of them, the conditions \eqref{eq:crital} and  \eqref{eq:critbe} define two critical lines, \(\alpha_c\) and \(\beta_c(\alpha)\) respectively, in the space of parameters \((\alpha,\beta)\). For \(\alpha_c<\alpha<1\) and \(\beta <\beta_c(\alpha)\), the system keeps spatially homogeneous. In this region, the predictions of Sec.~\ref{sec:approach} regarding the homogeneous approach to consensus, including the scaling of the distribution of opinions and Haff's law, are expected to be valid. For other values of the parameters, the system develops spatial non-homogeneities giving rise to potential novel approaches to consensus (strong non-homogeneous), not accounted for in Sec.~\ref{sec:approach}. This is investigated numerically in the next section. 

\section{Numerical simulations\label{sec:numsim}}

In this section we compare the theoretical results of Secs.~\ref{sec:approach} and \ref{sec:lattice} with Monte Carlo simulations. Other results, not addressed theoretically, are reported as well. The simulations were carried out following three different strategies, depending on whether the focus was on the entire temporal evolution of the system or on the final stage:
\begin{itemize}
\item[-] Whole time evolution. The system is simulated using the algorithm described in Sec.~\ref{sec:model} with:
  \begin{itemize}
  \item[+] \(\Delta s_t\) taken as constant. The global temperature \(T\) and \(\tau\), defined by Eq.~\eqref{eq:tiempotau}, as a function of time, are extracted from measuring the energy of the system, \(\sum\limits_{i\in\Sigma}s_i^2\),  and averaging over different realizations.
  \item[+] \(\Delta s_t\) taken as time dependent using a first realization, as explained at the in Sec.~\ref{sec:model}. The global temperature \(T\) is measured as before. 
  \end{itemize}
\item[-] Final stage. The algorithm in Sec.~\ref{sec:model} is slightly modified: \(\Delta s_t\) is taken as constant but after every social interaction all opinions are re-scaled to ensure that the total energy of the system keeps its initial value (as opposite of taken \(\Delta s_t\) from a first realization). In this way, the global temperature \(T\) remains always and exactly constant and the identification of a possible final stage is easier. Here, two main quantities are measured: the excess kurtosis
  \begin{equation}
    \label{eq:kurt}
    e=\frac{1}{N}\sum\limits_{i\in\Sigma}s_i^4-3
  \end{equation}
  and the opinion distribution. The former is measured using only one realization and is used to visualize the eventual transients and pre-asymptotic regimes, while the latter is averaged over all agents (or those of a given degree), all realizations, and over times in a time window located at the end of the trajectories.
\end{itemize}

Four different social interaction networks are studied: all-to-all, Erdös-Rényi, Barabási-Albert, and square 2D lattice. In each of the cases, the number of agents is fixed [\(N=100\) (all-to-all), \(500\) (Erdös-Rényi and Barabási-Albert), and \(30\times 30\) (square 2D lattice)] and the parameters \(\alpha\in(-1,1)\) and \(\beta\ge 0\) are changed  to cover a wide region of the parametric space. We have restricted the simulations to \(\beta \in[0,10]\) since for larger values of \(\beta\) the dynamics becomes very slow and the algorithm very inefficient. 

\subsection{All-to-all. Homogeneous approach  to consensus}

\begin{figure}[!h]
  \centering
  \includegraphics[angle=-90,width=.45\linewidth]{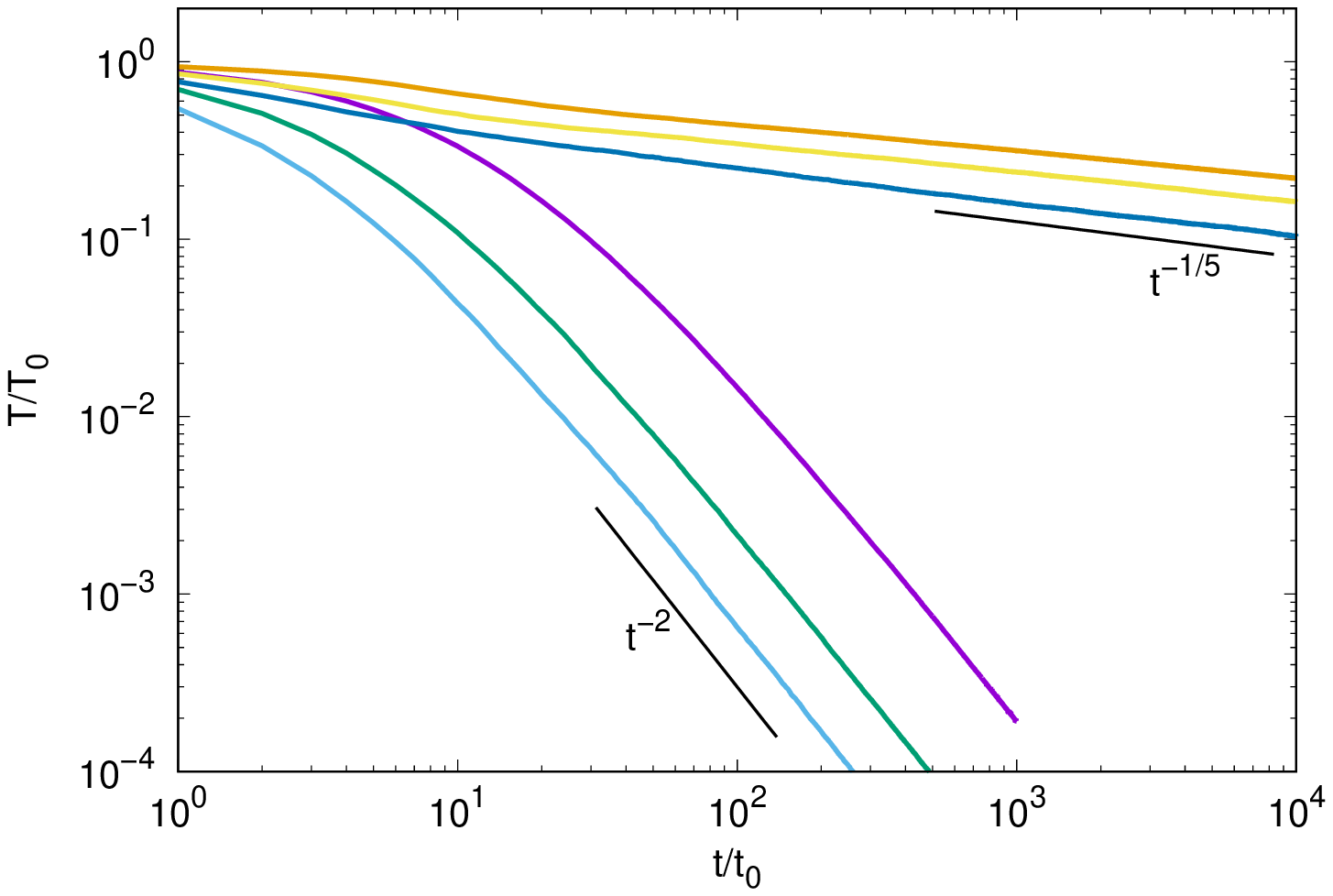}
  \hfill
  \includegraphics[angle=-90,width=.435\linewidth]{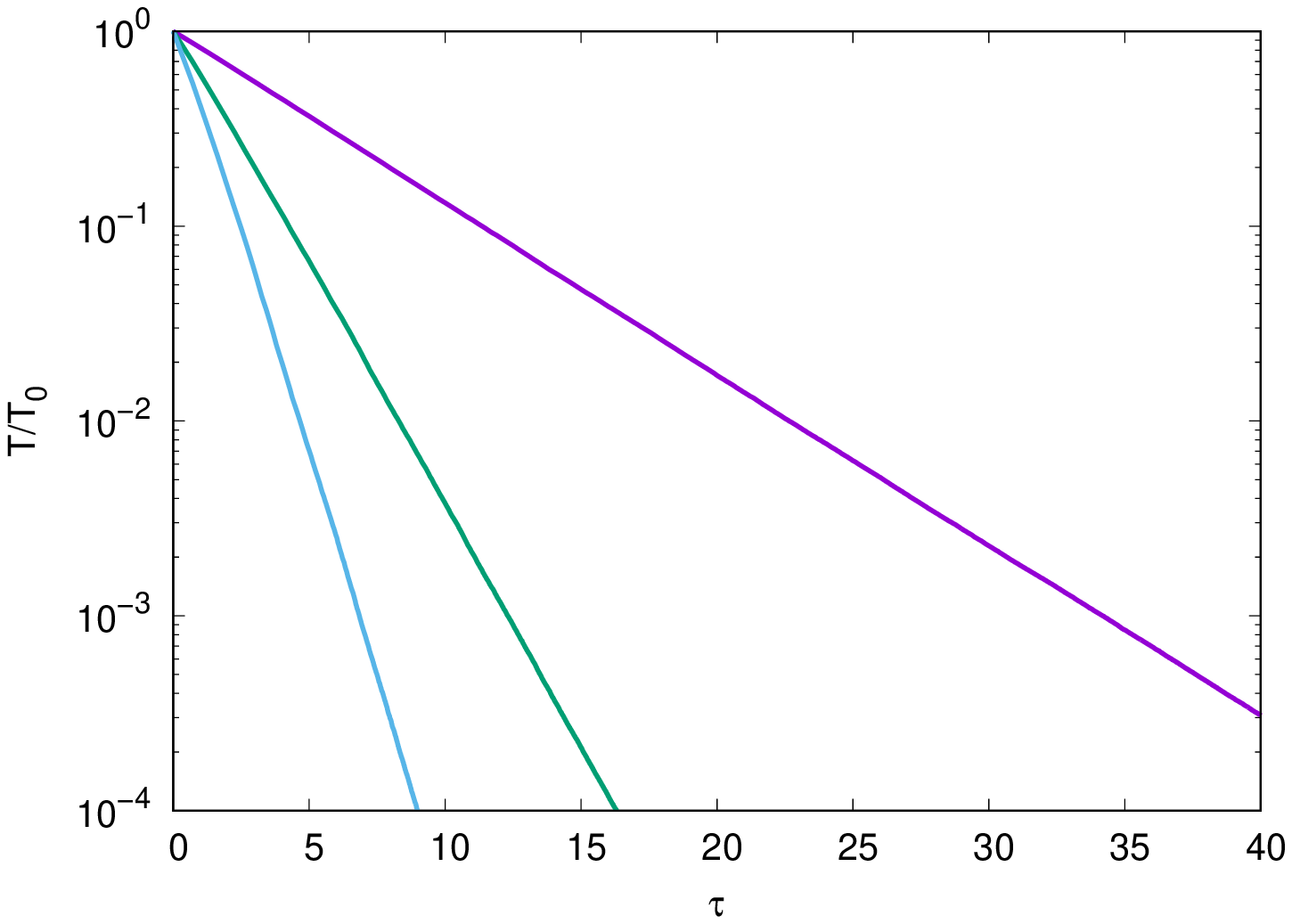} \\
  \includegraphics[angle=-90,width=.45\linewidth]{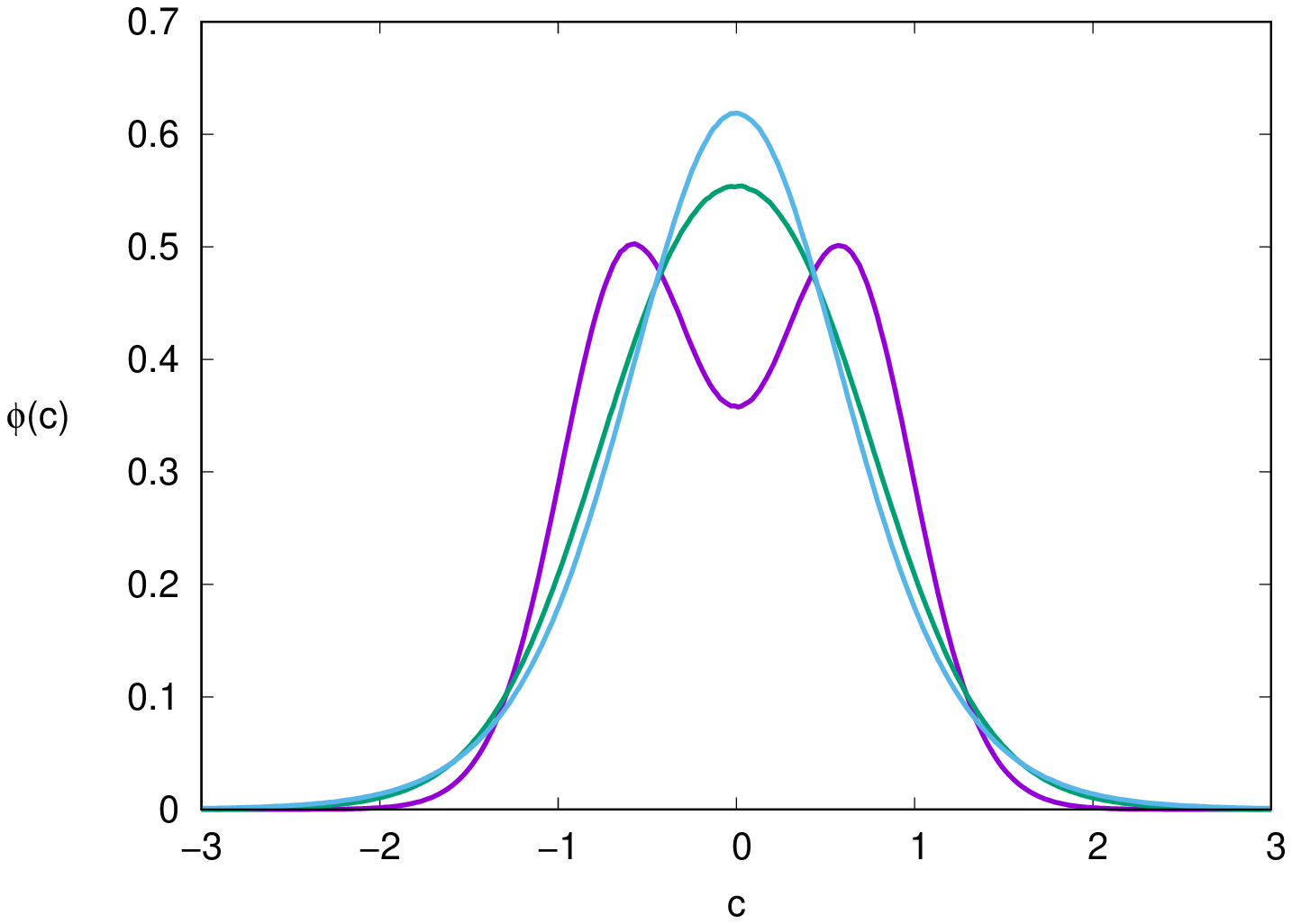}
  \hfill
  \includegraphics[angle=-90,width=.45\linewidth]{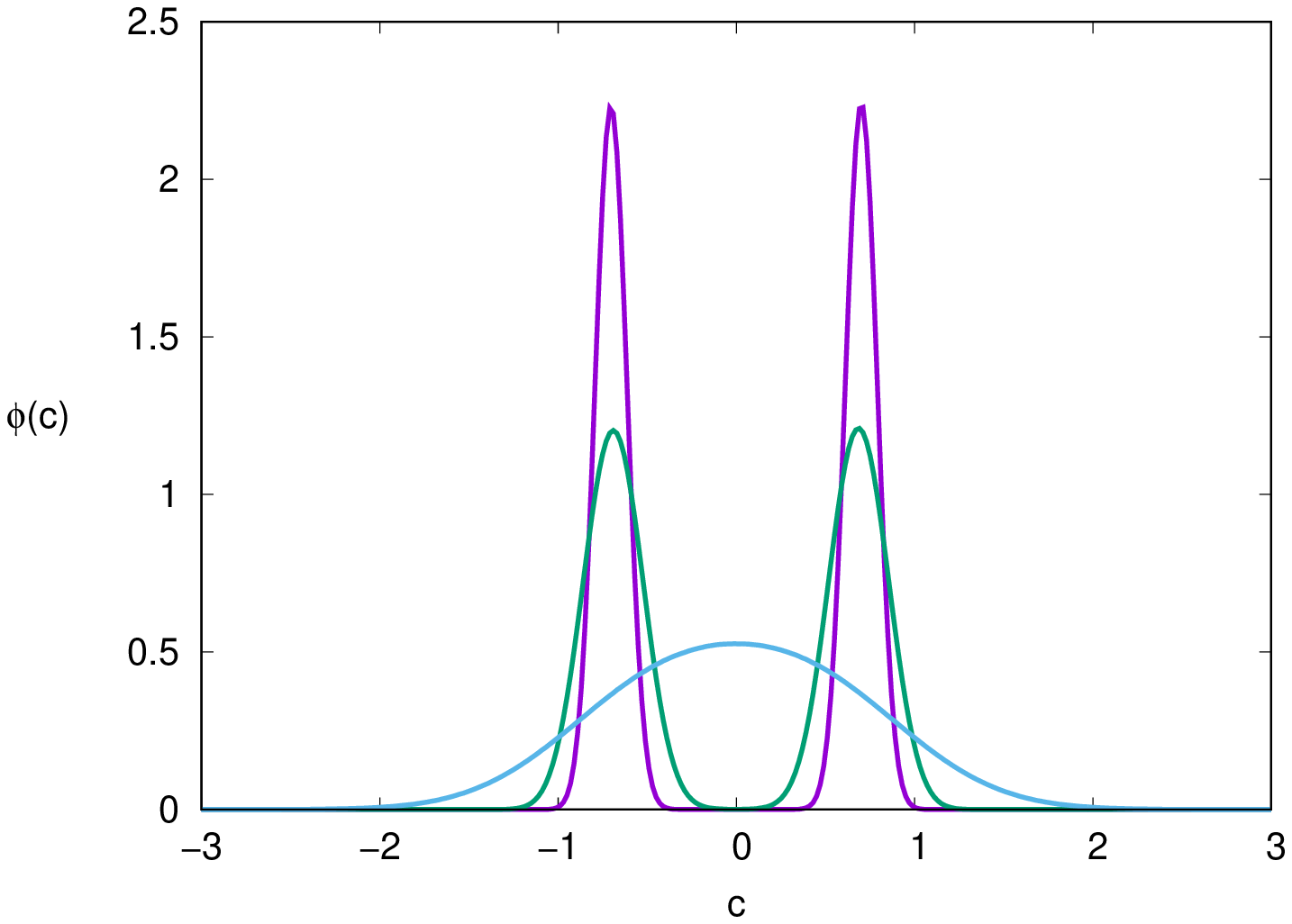}
  \caption{Simulations results of an all-to-all network with \(N=100\) nodes. Top-left plot: the global temperature \(T\) as a function of the time \(t\). For \(t/t_0=10^2\), from top to bottom, the lines correspond to \((\alpha,\beta)=(9/10,10),(7/10,10),(3/10,10),(9/10,1),(7/10,2),(3/10,1)\). The segments related to the decays \(t^{-1/5}\) and \(t^{-2}\) are the theoretical prediction of Eq.~\eqref{eq:ttbet>0} for \(\beta=10\) and \(\beta=1\), respectively. Top-right plot: the global temperature \(T\) as a function of the time \(\tau\) defined by Eq.~\eqref{eq:tiempotau}. From top to bottom: \((\alpha,\beta)=(9/10,1),(7/10,1),(3/10,1)\). Bottom-left plot: the scaled distribution \(\phi\) for \((\alpha,\beta)=(9/10,1),(7/10,1),(3/10,1)\). Bottom-right plot: the scaled distribution \(\phi\) for \((\alpha,\beta)=(9/10,10),(7/10,10),(3/10,10)\). }
  \label{fig:1}
\end{figure}

Here we present our results for a fully connected social interaction network. In this case, all agents interact with each other, and a homogeneous approach to consensus, as described in Sec.~\ref{sec:approach}, is observed.

The time evolution of the global temperature \(T\) is shown in the top plots of Fig.~\ref{fig:1}. In agreement with the Haff's law, \(T(t)\) tends to a power law (left plot) with the exponent depending only on \(\beta\), according to Eqs.~\eqref{eq:ttbet0} and \eqref{eq:ttbet>0}, while \(T(\tau)\) follows an exponential (right plot), Eq.~\eqref{eq:ttexp}. Interestingly, the results depend only on \(|\alpha|\) and \(\beta\), the sign of \(\alpha\) being irrelevant. 

The bottom plots in Fig.~\ref{fig:1} show the scaled opinion distribution \(\phi\), defined by Eq.~\eqref{eq:scalpij}, for some values of \(\alpha\) and \(\beta\). As discussed in Ref.~\cite{khalil2021approach}, for a given value of \(|\alpha|\), below a critical value of \(\beta(|\alpha|)\), the scaled distribution \(\phi\) is unimodal, and above it, bimodal.

The symmetry between \(\alpha\) and \(-\alpha\) shown by the numerical simulations is due to the possibility of interpreting a social interaction with \(\alpha\) as one with \(-\alpha\) followed by a permutation of the agents' locations (nodes). Since in the all-to-all setting, locations are not important, social interactions with \(\alpha\) and \(-\alpha\) are equivalent, as already analyzed in Ref.~\cite{khalil2021approach}. This symmetry is lost in more structured interactions, as we see next. 

\subsection{Erdös-Rényi and Barabási-Albert. Weak non-homogeneous approach to consensus}

\begin{figure}[!t]
  \centering
  \includegraphics[angle=-90,width=.45\linewidth]{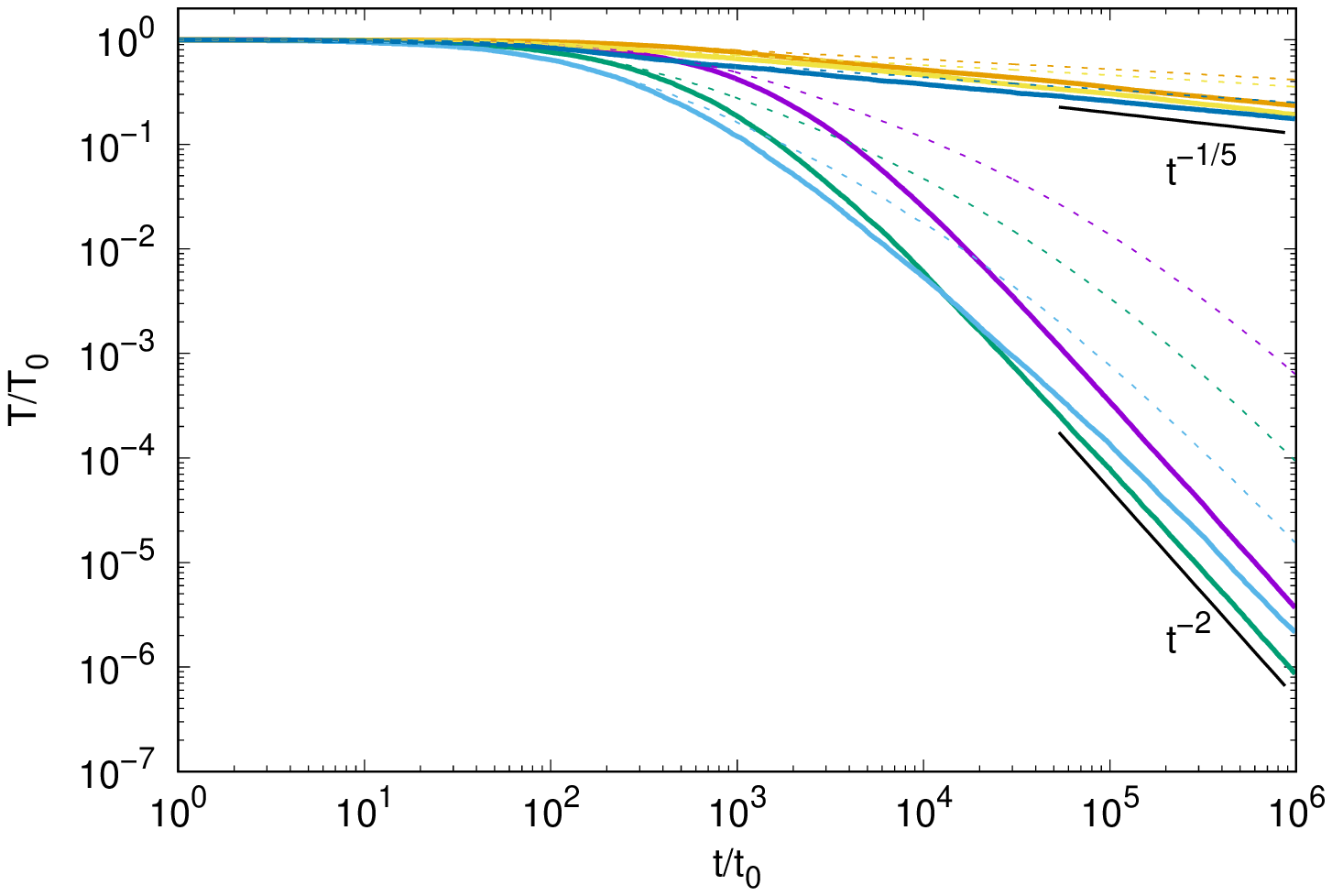}
  \hfill
  \includegraphics[angle=-90,width=.435\linewidth]{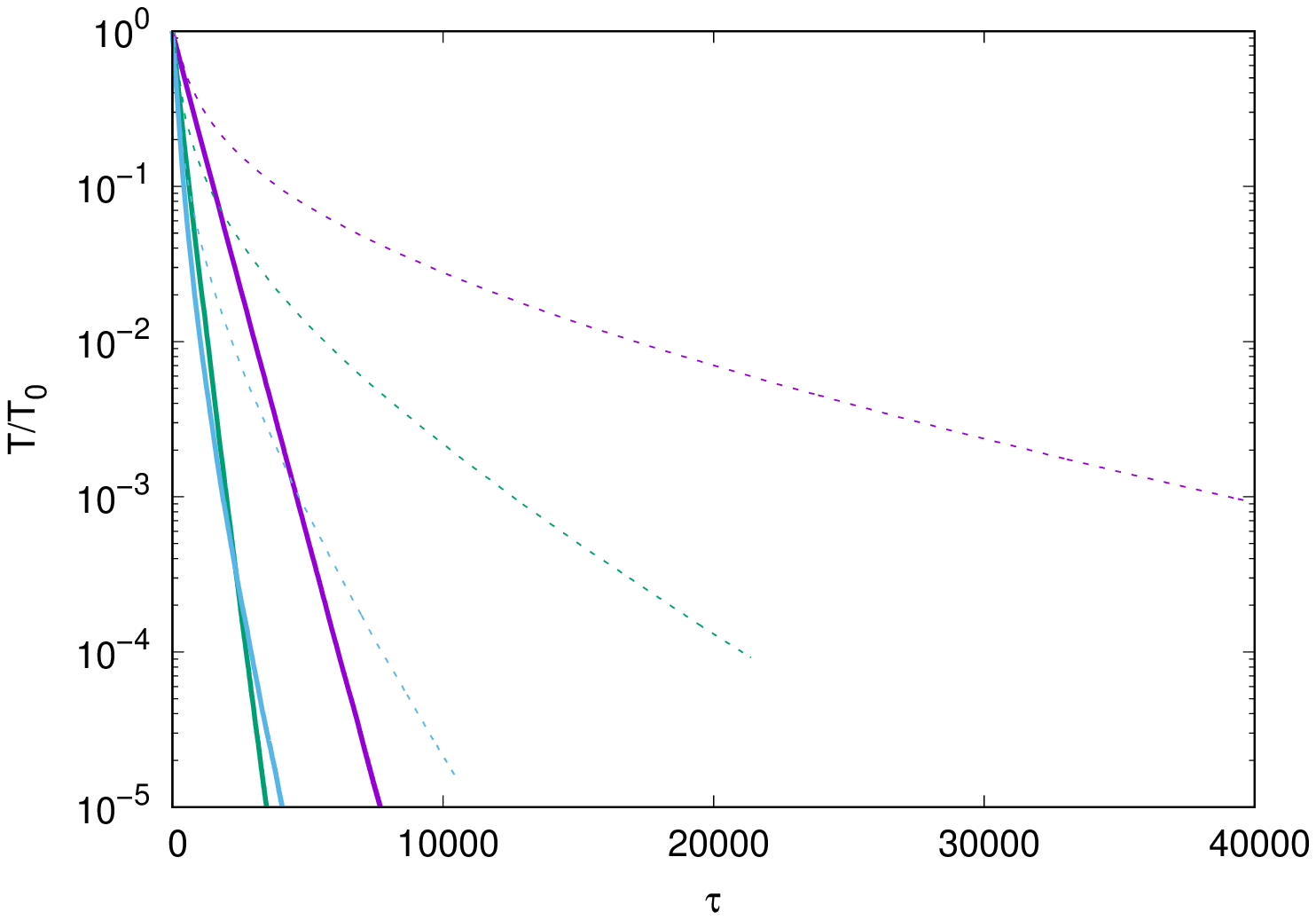} \\
  \includegraphics[angle=-90,width=.45\linewidth]{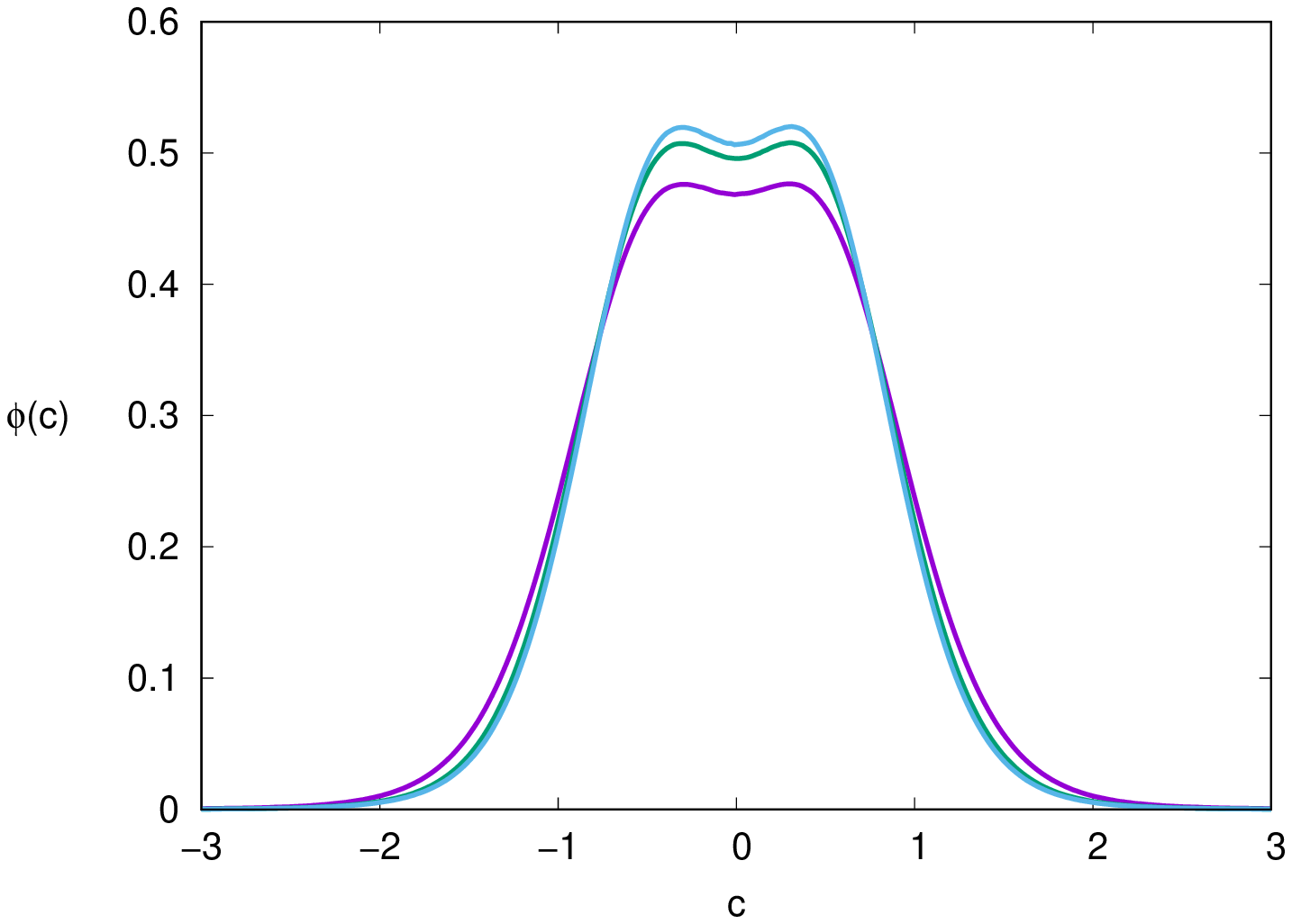}
  \hfill
  \includegraphics[angle=-90,width=.45\linewidth]{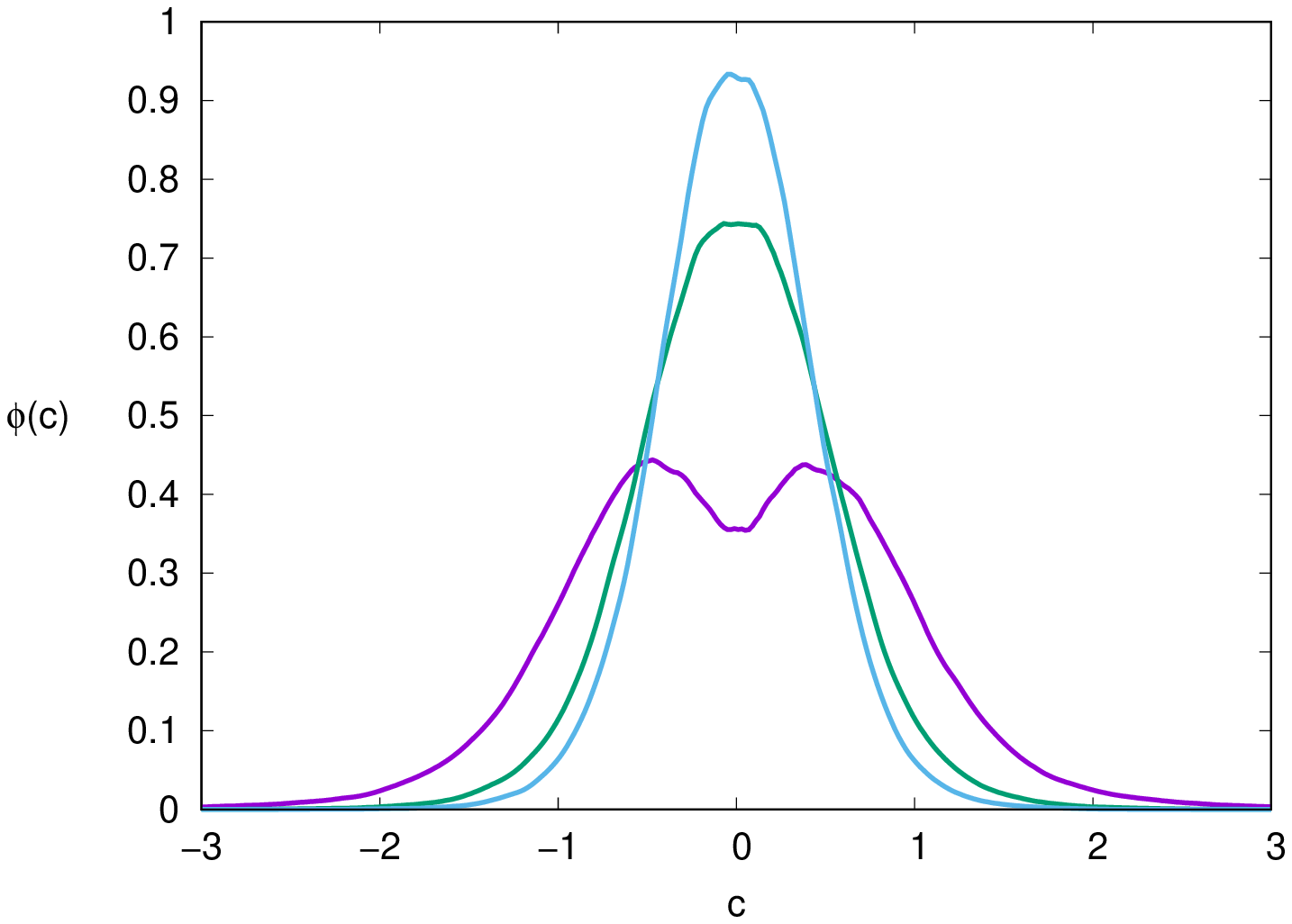}
  \caption{Simulations results for an Erdös-Rényi network with \(N=500\) agents and mean degree \(4\). The top plots are analogous to Fig.~\ref{fig:1}, with additional dashed lines that correspond to \(-\alpha\). The bottom-left plot is as the corresponding one in Fig.~\ref{fig:1}. The bottom-right plot shows the scaled distribution function \(\phi\) for the sub-group of agents with degree \(2\) (bimodal distribution), degree \(4\), and \(6\) (unimodal ones).}
  \label{fig:2}
\end{figure}

\begin{figure}[!h]
  \centering
  \includegraphics[angle=-90,width=.45\linewidth]{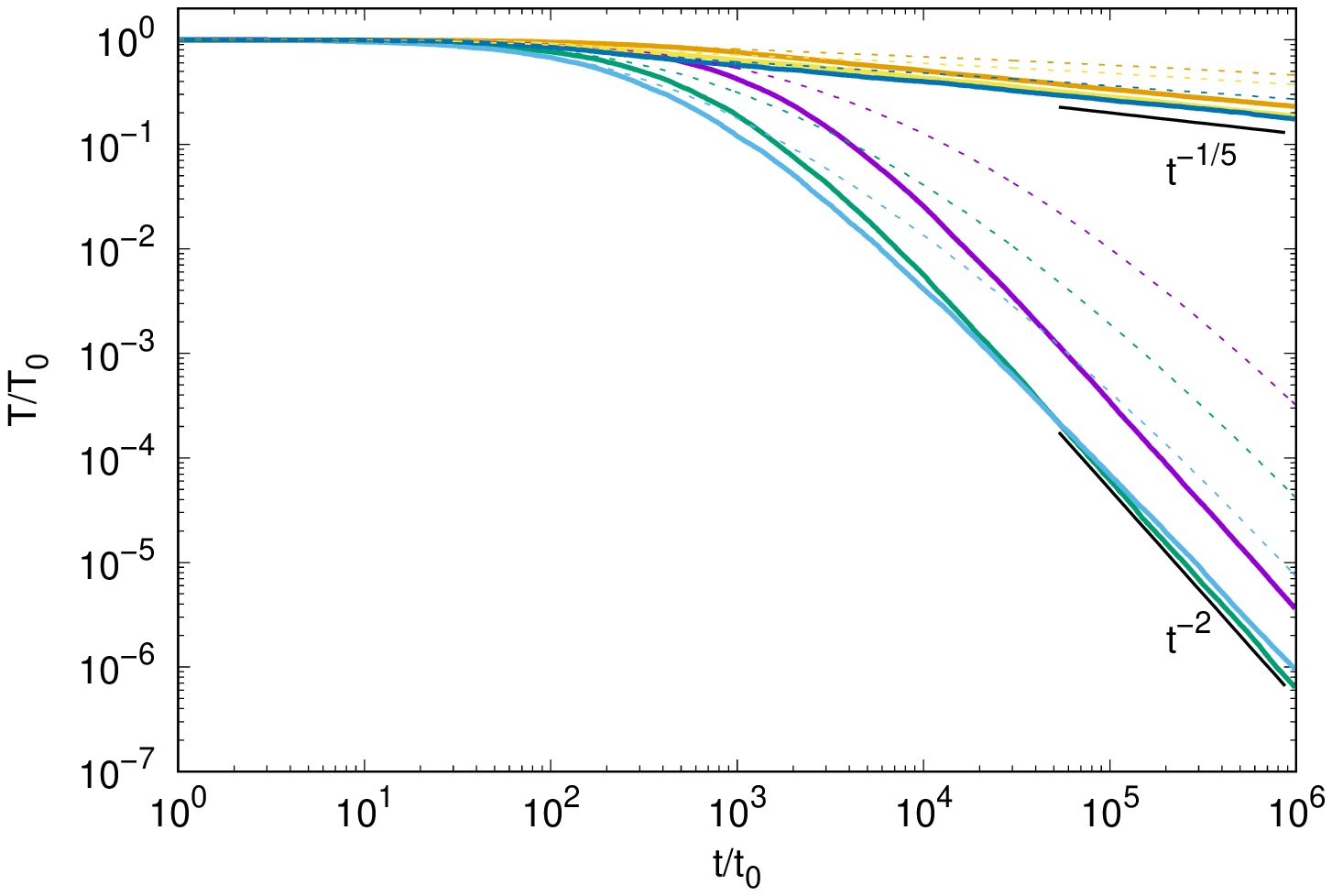}
  \hfill
  \includegraphics[angle=-90,width=.435\linewidth]{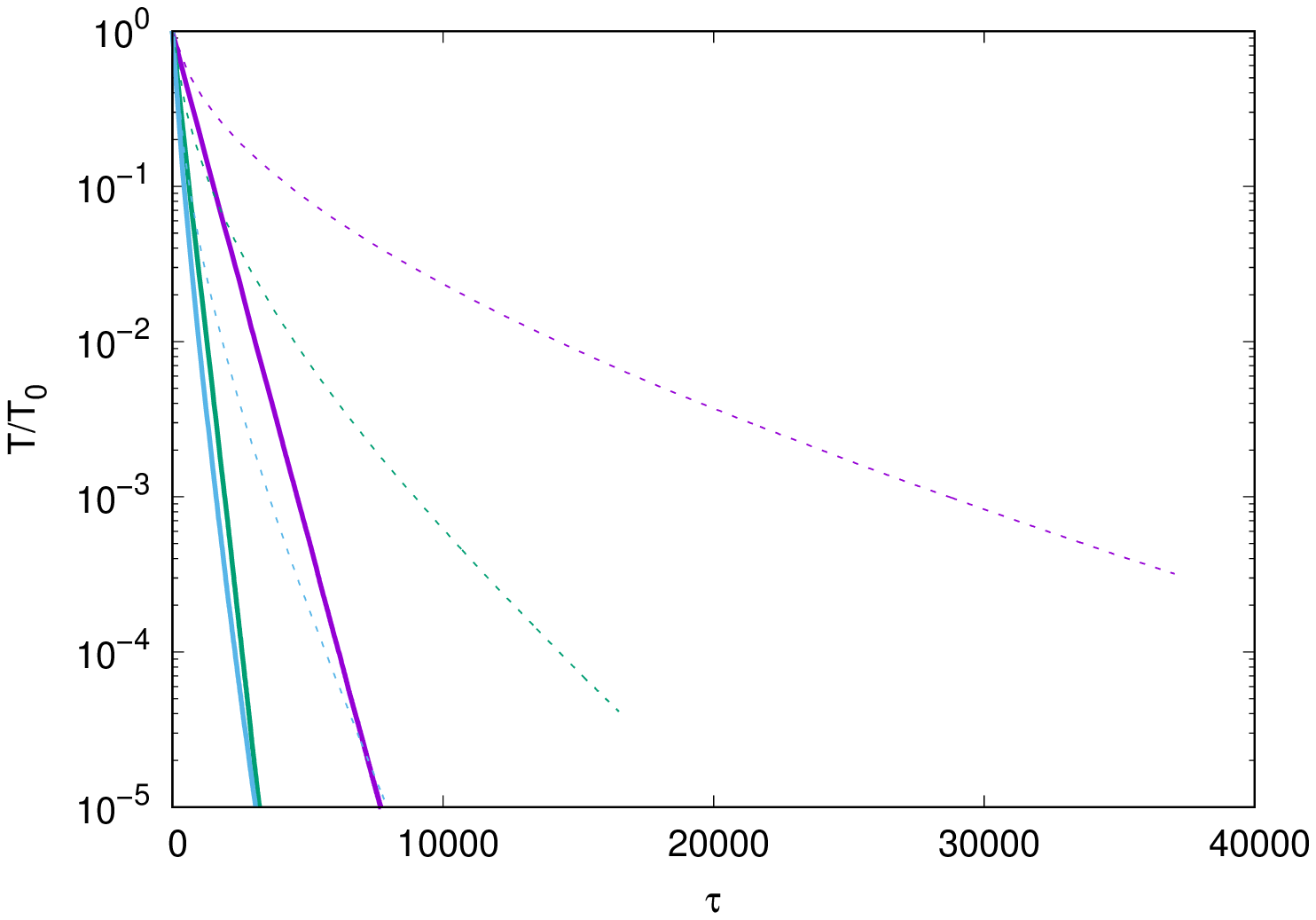}\\
  \includegraphics[angle=-90,width=.45\linewidth]{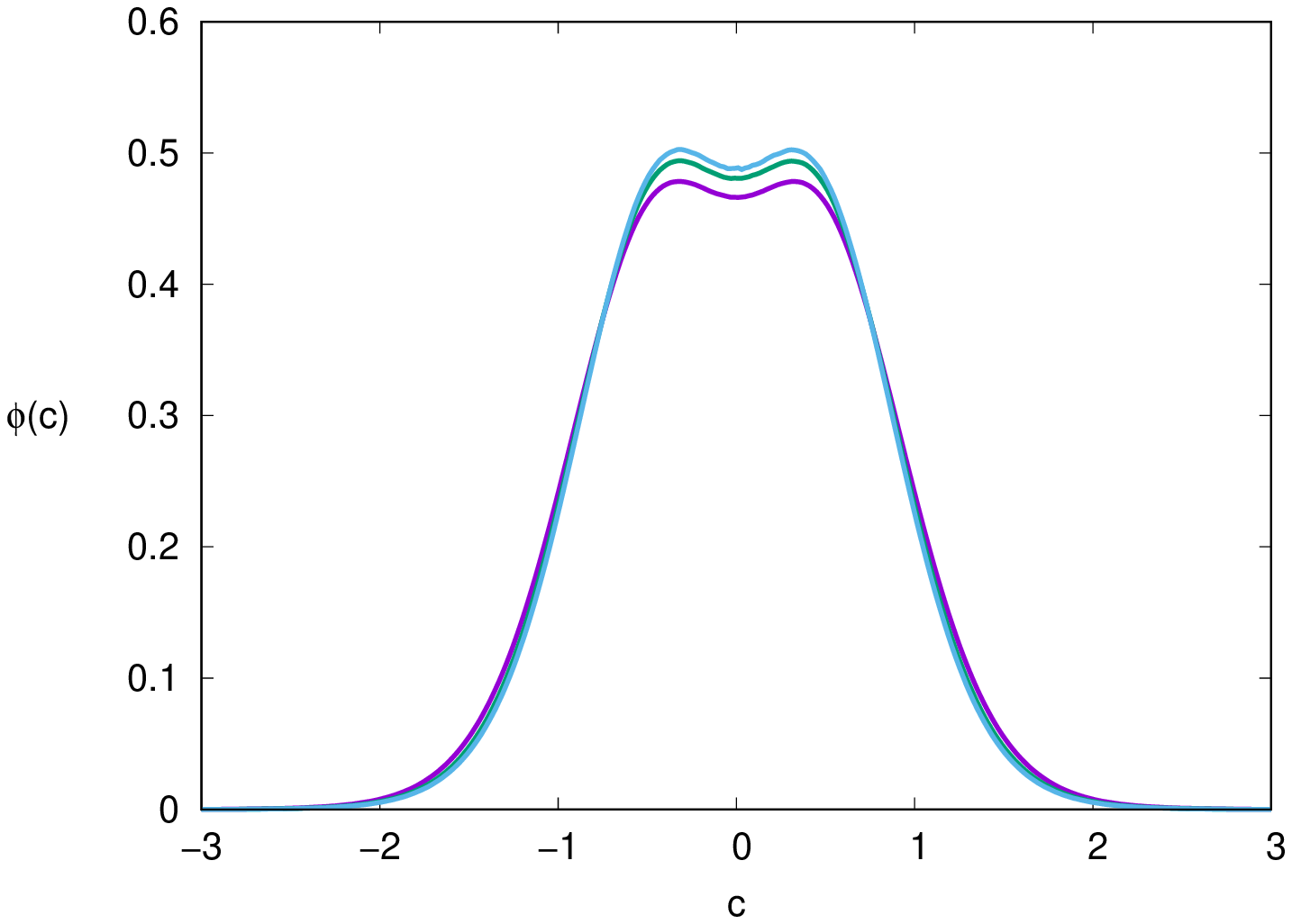}
  \hfill
  \includegraphics[angle=-90,width=.45\linewidth]{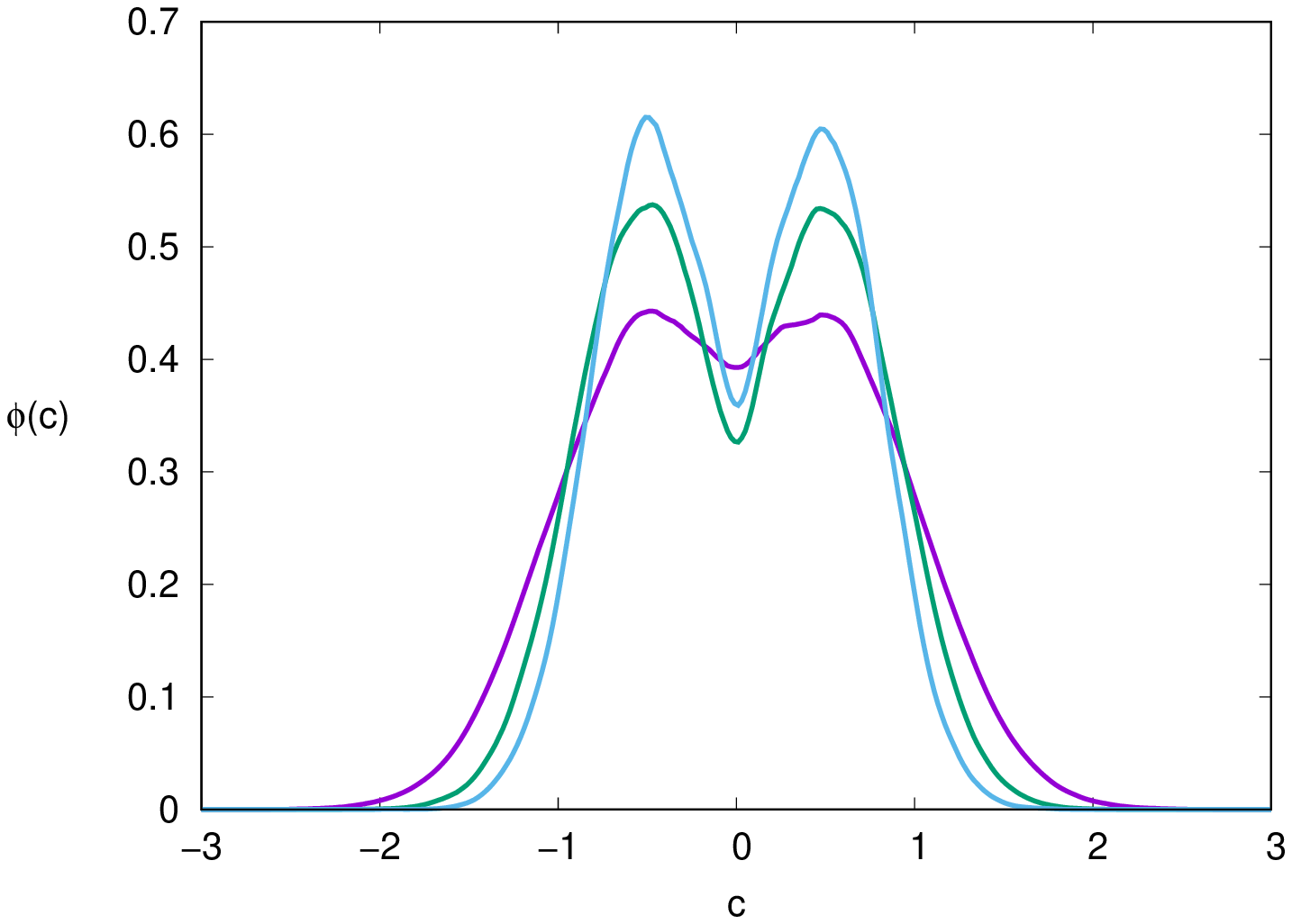}
  \caption{Simulations results for a Barabási-Albert network with \(N=500\) agents and mean degree \(4\). The plots show the same data as in Fig.~\ref{fig:2}.}
  \label{fig:3}
\end{figure}

The results of the Erdös-Rényi and Barabási-Albert networks, with mean degree \(4\), are illustrated in the Figs.~\ref{fig:2} and \ref{fig:3}, respectively. In both cases, the results are compatible with a weak non-homogeneous approach to consensus characterized by Eqs.~\eqref{eq:nohom1} and \eqref{eq:nohom2}.

In both networks, we observe a similar behaviour. After a transient period, which can be very long, the mean opinions become zero and all local temperatures divided by the global one, as well as the scaled opinion distribution, become time-independent. Furthermore, in the long run, each agent has a different distribution of opinions and therefore a different temperature, in general. The results for \(\alpha\) and \(-\alpha\) are clearly different.

The upper plots in Figs.~\ref{fig:2} and \ref{fig:3} show the global temperature \(T\) as a function of time \(t\) (left plot) and time \(\tau\) (right plot) for the same values of \(\alpha\) and \(\beta\) as in Fig.~\ref{fig:1}. The long-lime behaviour of \(T(t)\) follows Haff's law. However, for a given value of \(\beta\), the relaxation of the system for \(\alpha<0\) is slower than for \(-\alpha>0\). The latter is much more evident looking at \(T(\tau)\) for \(\alpha<0\) which experiences significant initial deviation from the exponential decay, a clear indication of a transient period. 

The bottom plots in Fig.~\ref{fig:2} and \ref{fig:3} provide the scaled opinion distribution for nodes of different degrees, and for different values of \(\alpha\) and \(\beta\). It is apparent that different groups of nodes with different degrees have different distributions and temperatures. Moreover, in the same network and for some values of the parameters, agents with higher degree have unimodal distributions while that with lower degree have bimodal distributions.

\subsection{Square 2D lattice. Phase diagram}

\begin{figure}[!h]
  \centering
  \includegraphics[width=.75\linewidth]{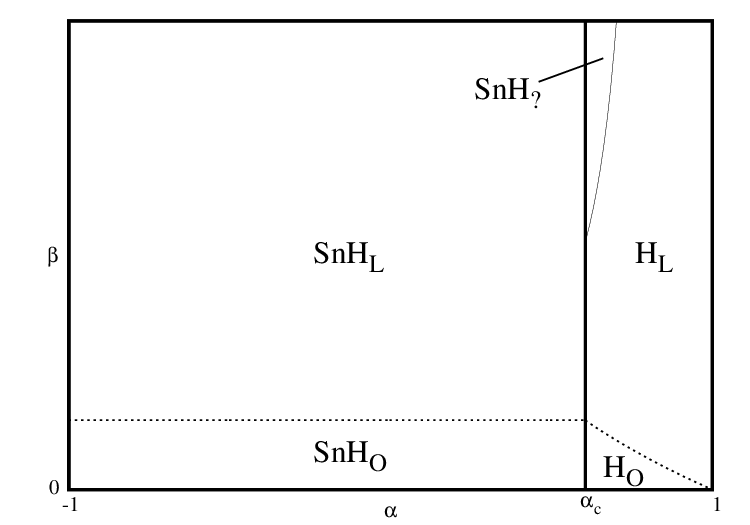}
  \caption{Sketch of the phase diagram with the five configurations described in the main text; see also next Figures. The vertical line at \(\alpha_c\) is given by the critical condition \eqref{eq:crital} of the linear stability analysis of Sec.~\ref{sec:lattice}. The thin line, meeting the previous one, is given by the other condition \eqref{eq:critbe}. The dashed lines indicate smooth/continuous transitions from one configurations to the others. }
  \label{fig:4}
\end{figure}

The behaviour of the system on a 2D square lattice is richer than in previous cases, mainly because the system becomes spatially inhomogeneous for some parameter values, as anticipated in Sec.~\ref{sec:lattice}. To get an overview of the different scenarios, we first show in Fig.~\ref{fig:4} a phase diagram with different spatial opinion configurations that the system reaches after a long simulation (of the order of \(10^6t_0\)) as the parameters \(\alpha\) and \(\beta\) change.

For a given number of agents \(N\), the space of parameters \((\alpha,\beta)\) can be divided into five regions, as sketched in Fig.~\ref{fig:4}. The exact sizes of the regions depend on the actual value of \(N\).

\begin{figure}[!h]
  \centering
  \includegraphics[angle=-90,width=.45\linewidth]{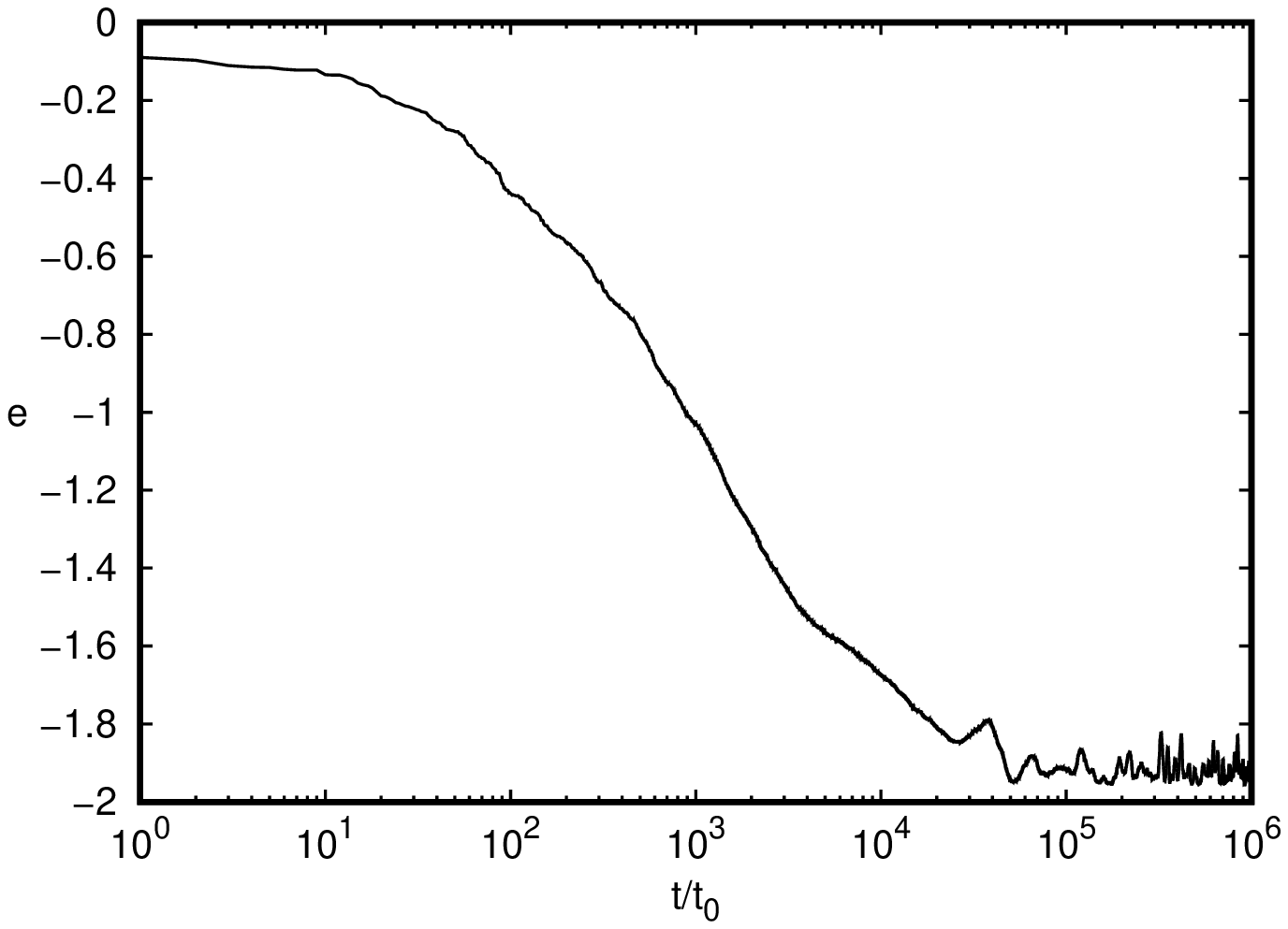}
  \hfill
  \includegraphics[angle=-90,width=.45\linewidth]{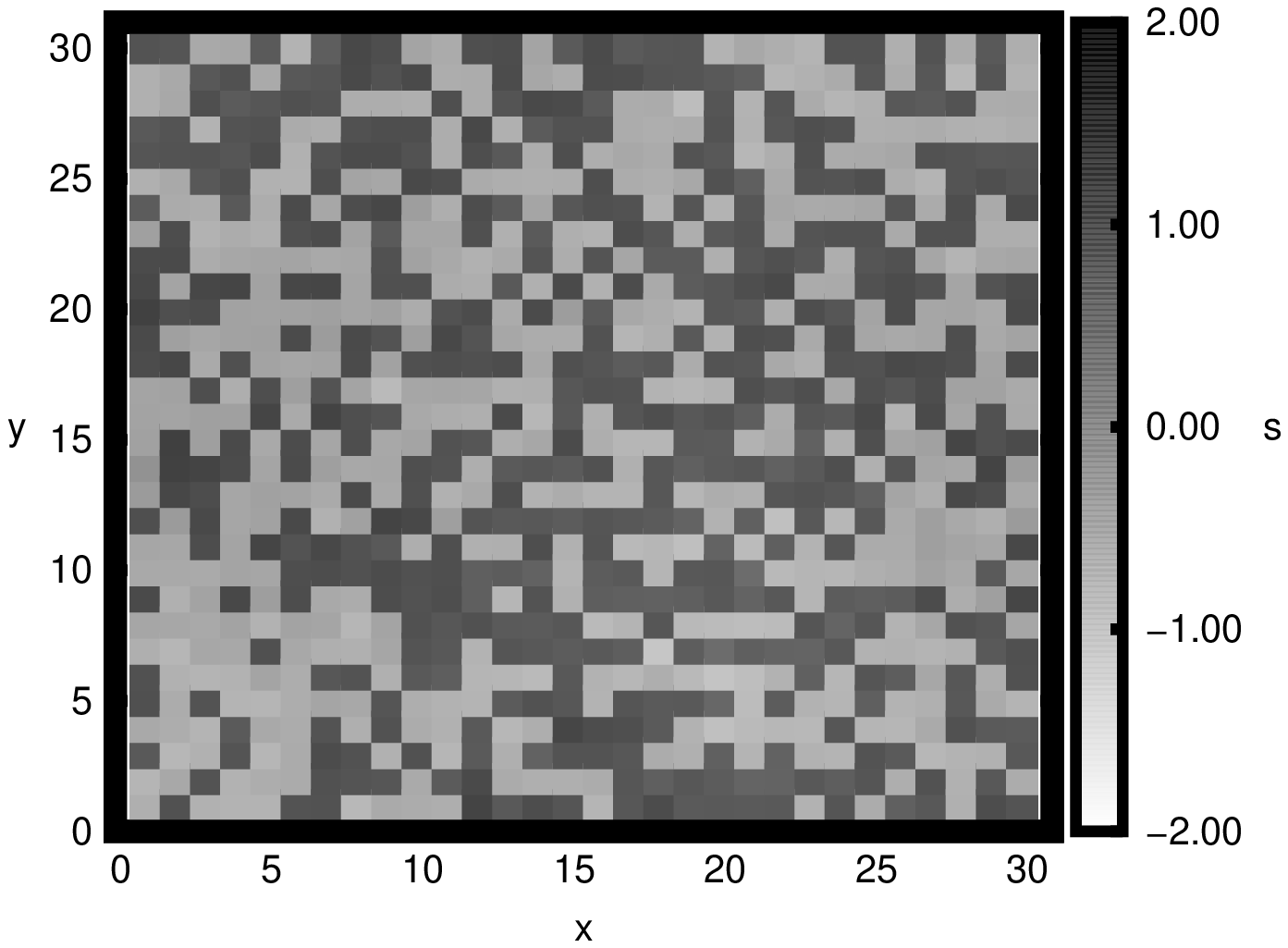} \\
  \includegraphics[angle=-90,width=.45\linewidth]{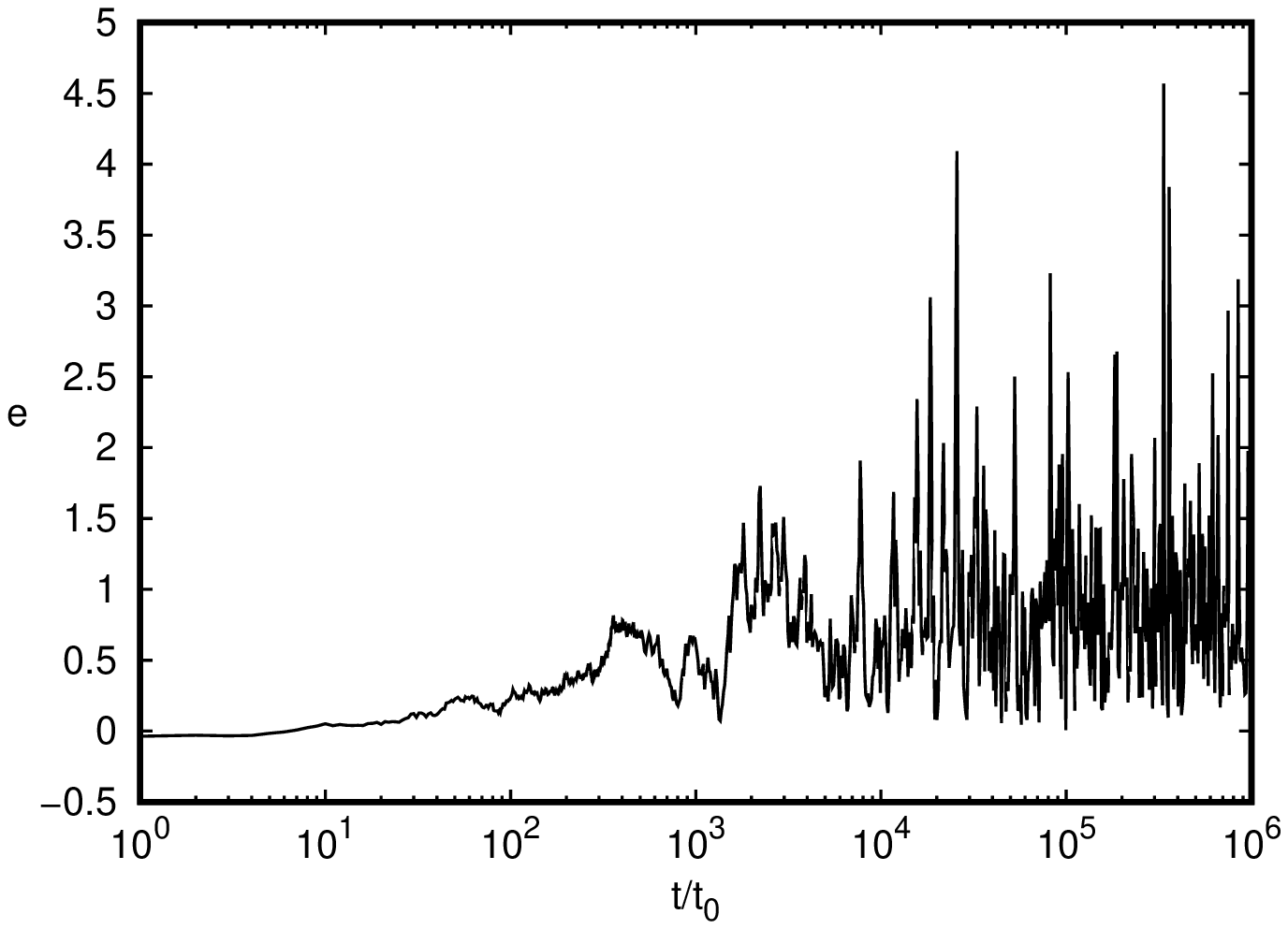}
  \hfill
  \includegraphics[angle=-90,width=.45\linewidth]{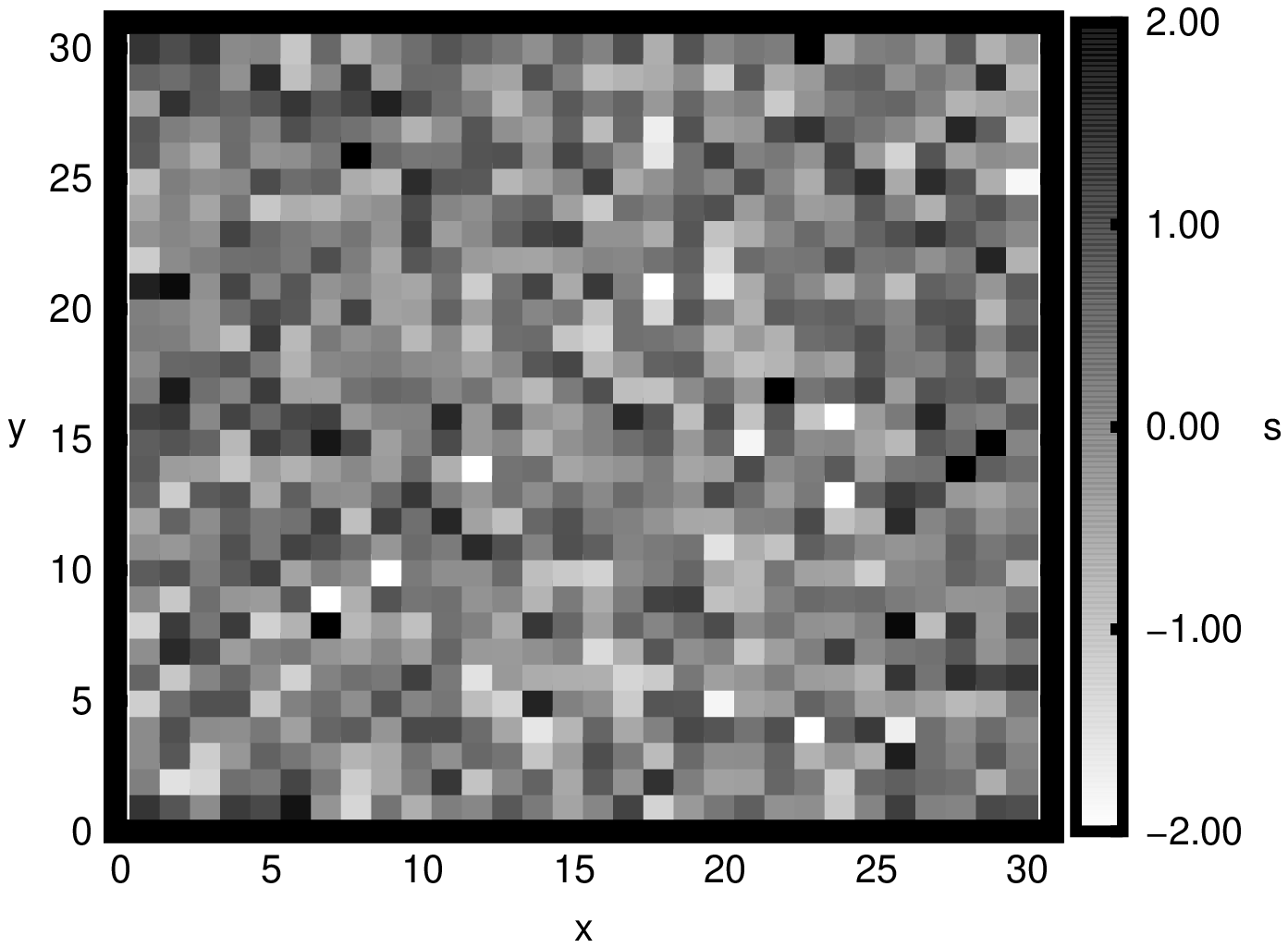}
  \caption{Simulation results of a system of \(N=900\) agents on a square 2D lattice with \(\alpha=0.98\). The left column shows the time evolution of the kurtosis \(e\), defined by Eq.~\eqref{eq:kurt}, and the right column the final state of the system. Bottom: \(\beta=0\), in H\(_\text{O}\). Top: \(\beta=5\), in H\(_\text{L}\). }
  \label{fig:5}
\end{figure}

In the H\(_\text{O}\) region, the system states spatially homogeneous, as can be seen in the bottom-right plot of Fig.~\ref{fig:5} for \(\alpha=0.98\) and \(\beta=0\). The evolution of the kurtosis (bottom-right plot) indicates a transient of the order of \(10^3t_0\): a first monotonic evolution followed by a very noisy plateau. The scaled distribution function of the system is unimodal (not shown).

From H\(_\text{O}\), upon increasing \(\beta\), the system keeps spatially homogeneous, but the local structure changes. For \(\beta\) big enough, we identify the H\(_\text{L}\) region: agents tend to arrange forming filaments of L form and similar opinions, homogeneously distributed, as shown in the top-right plot of Fig.~\ref{fig:5}. The evolution of the kurtosis shows a longer transient after a plateau stating around \(10^5t_0\). In H\(_\text{L}\), the scaled distribution function is bimodal (not shown).

The transition from H\(_\text{O}\) to H\(_\text{L}\) is smooth and approximately given by the unimodal-bimodal line obtained in Ref.~\cite{khalil2021approach} (the dashed line at the bottom-right of Fig.~\ref{fig:4}). As \(\beta\) increases from the H\(_\text{O}\) region, the presence of the L local structures becomes more evident. Moreover, for larger \(\beta\) the population of agents is clearly polarized into two groups of opposite opinions. Only a small fraction of the total agents have moderate opinions: this is quite apparent from the bimodal form of the scaled distribution function which is similar to the ones in the bottom-right plot of Fig.~\ref{fig:1}. The L structures are not static, but are created, destroyed, and move with time.

Starting from H\(_\text{L}\), by further increasing \(\beta\), we can reach the theoretical region SnH\(_\text{?}\), see Fig.~\ref{fig:4}. According to the theoretical results of Sec.~\ref{sec:lattice}, SnH\(_\text{?}\) corresponds to spatially inhomogeneous solutions of the hydrodynamic equations originating from the linear instability of the temperature field, Eq.~\eqref{eq:critbe}. However, for not too small values of \(N\), the SnH\(_\text{?}\) region is found at large values of \(\beta\) that simulations cannot explore; the dynamics become too slow and everything observed seems to be part of a transient.

From the regions H\(_\text{O}\) and H\(_\text{L}\), as \(\alpha\) decreases, the system becomes spatially unstable. In this way, we identify two new regions: SnH\(_\text{O}\) and SnH\(_\text{L}\). The transition from stable to unstable occurs for a specific critical value \(\alpha=\alpha_c\), expressed with great precision by Eq.~\eqref{eq:crital} and represented in Fig.~\ref{fig:4} by a thick vertical line. However, the transition between SnH\(_\text{O}\) and SnH\(_\text{L}\) is smooth. As a guide for the eye, it has been represented in Fig.~\ref{fig:4} by a horizontal dashed line. 

\begin{figure}[!h]
  \centering
  \includegraphics[angle=-90,width=.45\linewidth]{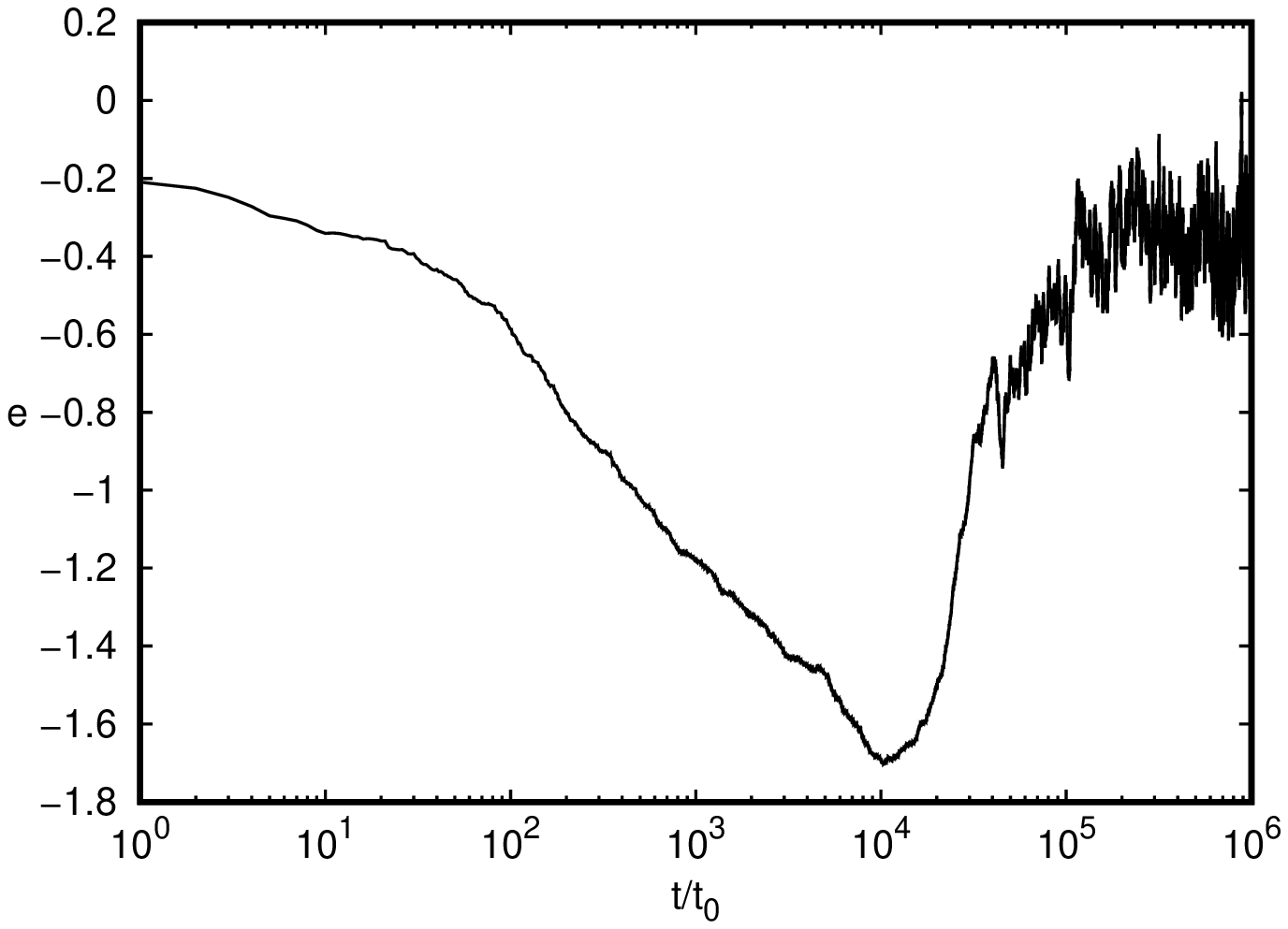}
  \hfill
  \includegraphics[angle=-90,width=.45\linewidth]{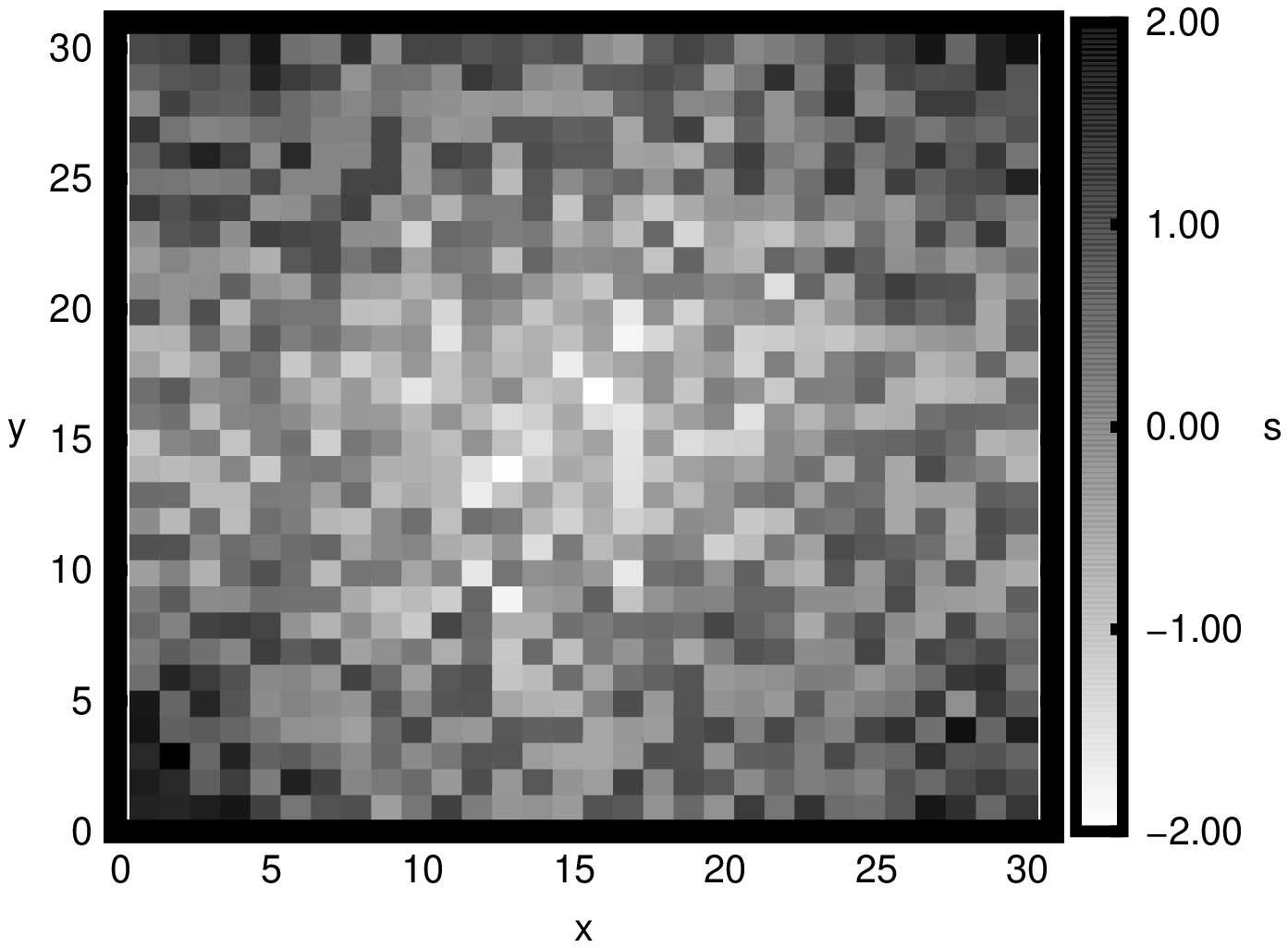} \\
  \includegraphics[angle=-90,width=.45\linewidth]{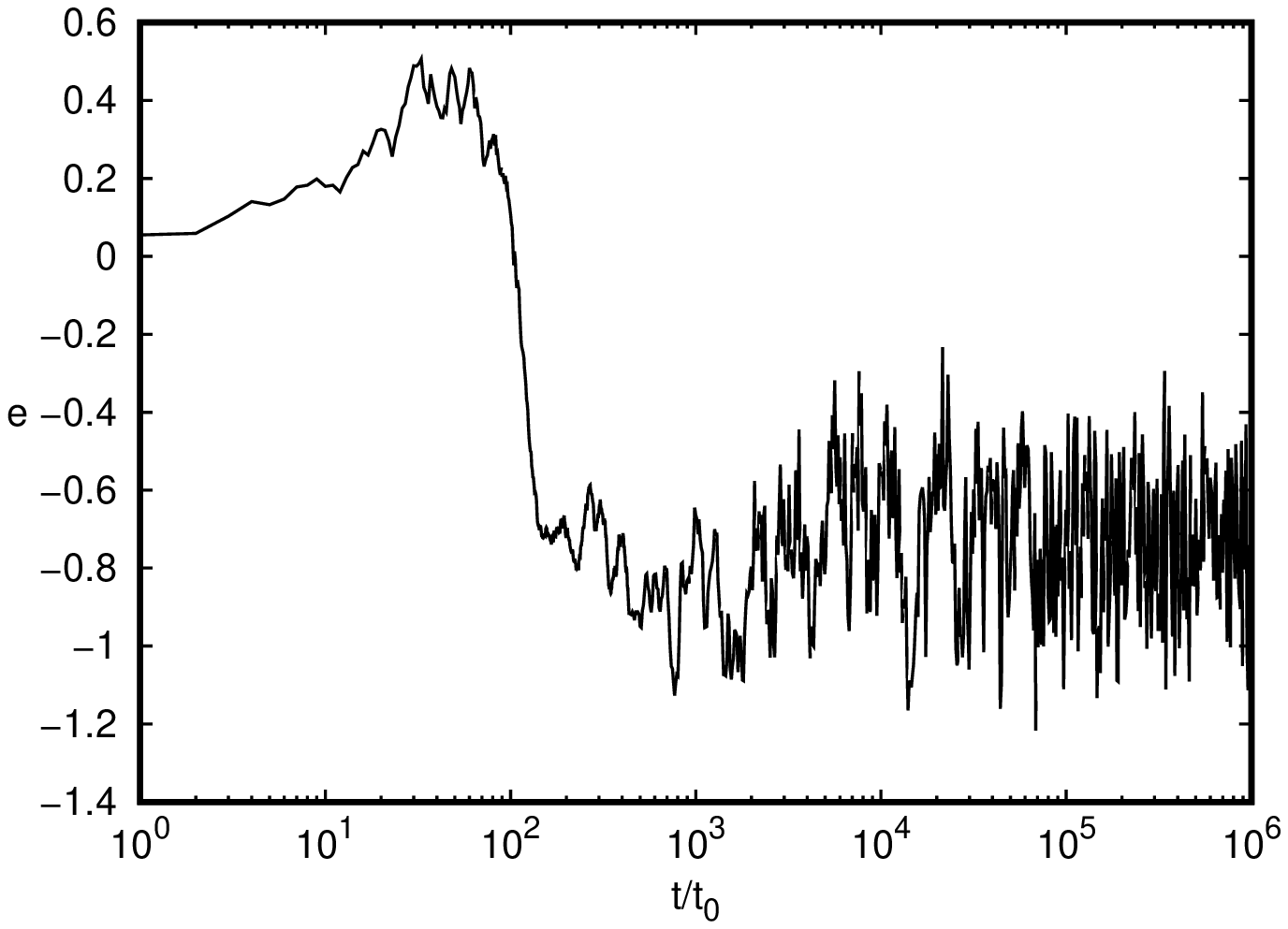}
  \hfill
  \includegraphics[angle=-90,width=.45\linewidth]{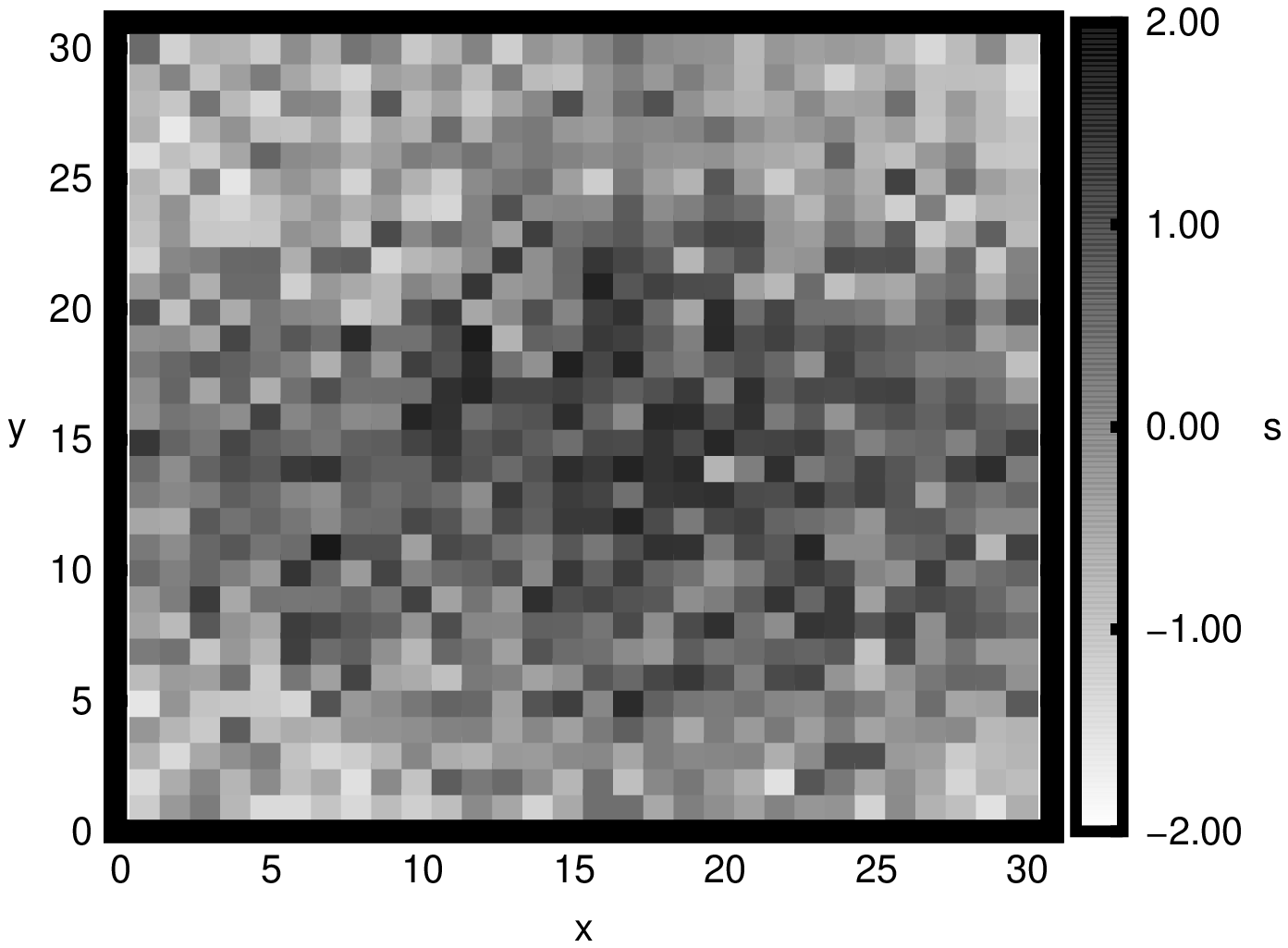}
  \caption{All plots as in Fig.~\ref{fig:5} but with \(\alpha=0.96\) and regions SnH\(_\text{O}\) (bottom) and SnH\(_\text{L}\) (top).}
  \label{fig:6}
\end{figure}

\begin{figure}[!h]
  \centering
  \includegraphics[angle=-90,width=.45\linewidth]{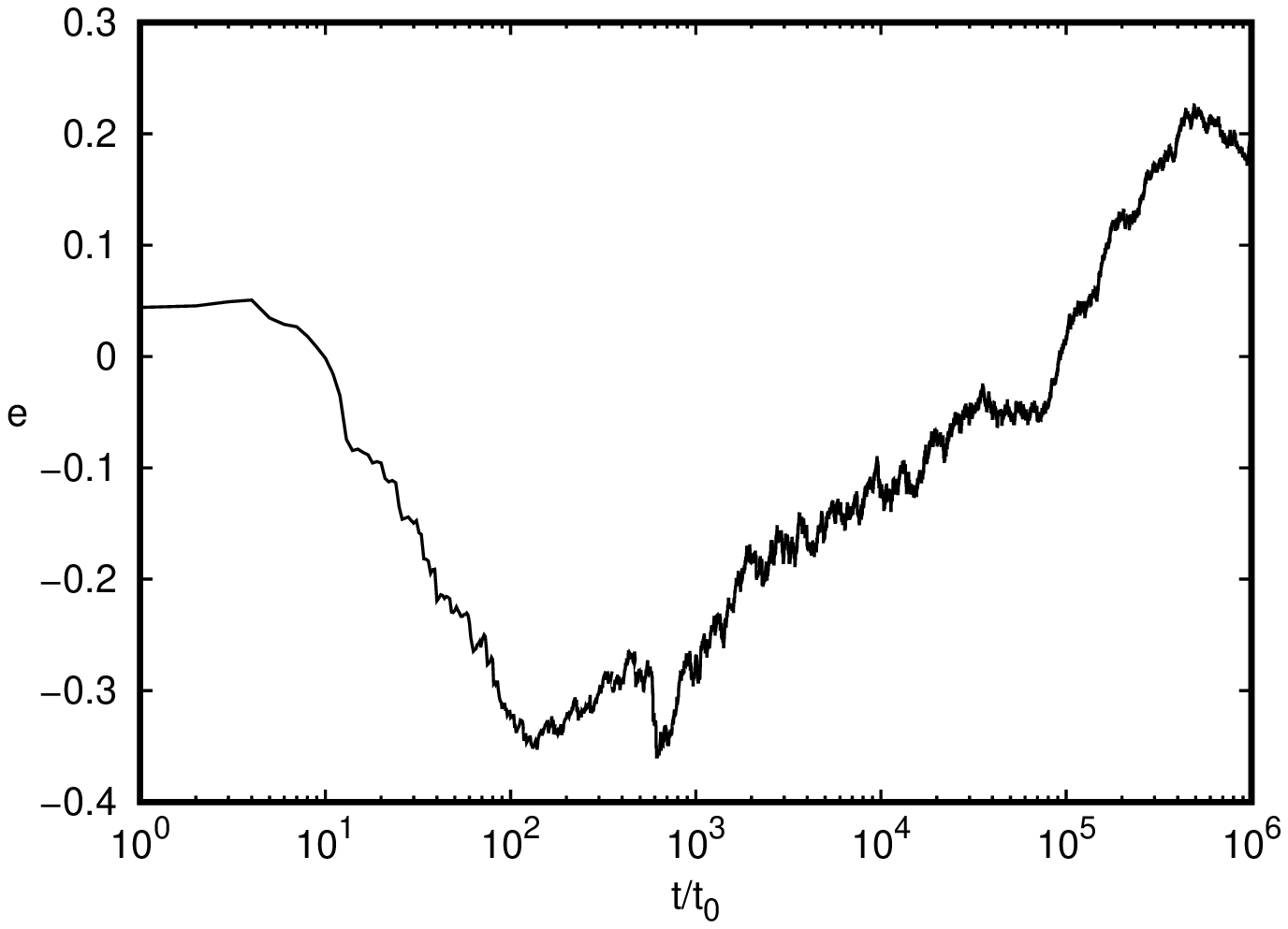}
  \hfill
  \includegraphics[angle=-90,width=.45\linewidth]{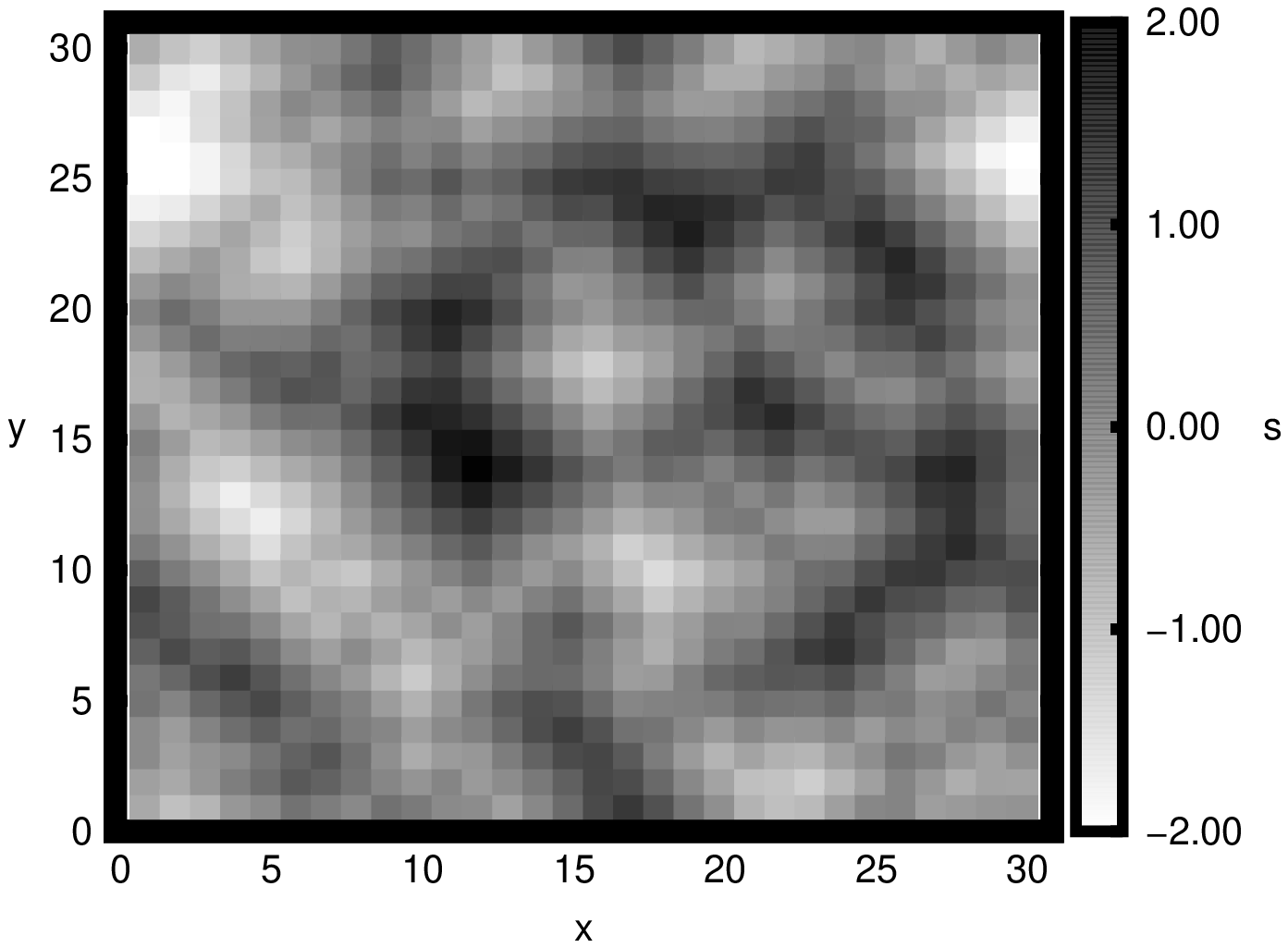} \\
  \includegraphics[angle=-90,width=.45\linewidth]{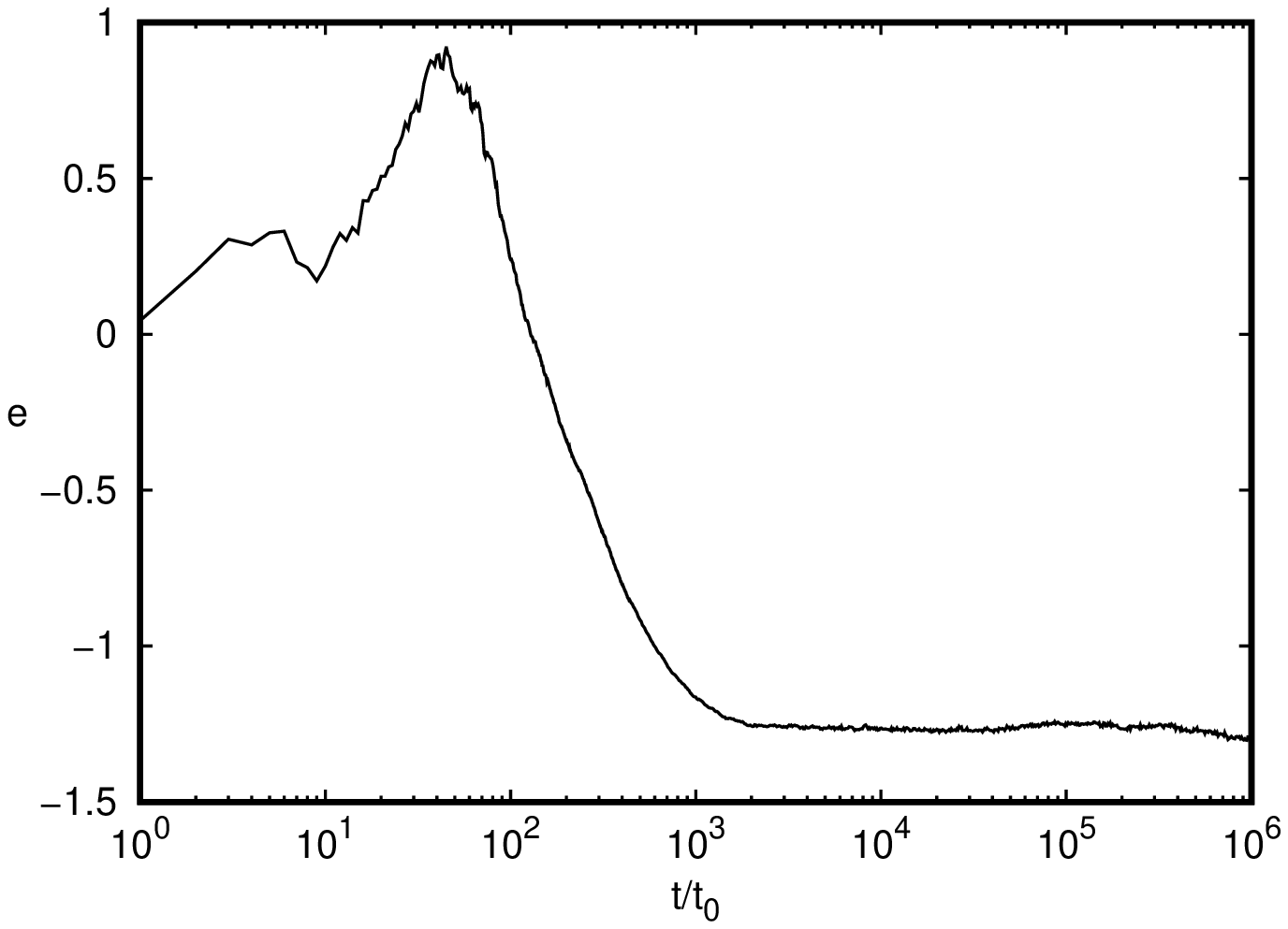}
  \hfill
  \includegraphics[angle=-90,width=.45\linewidth]{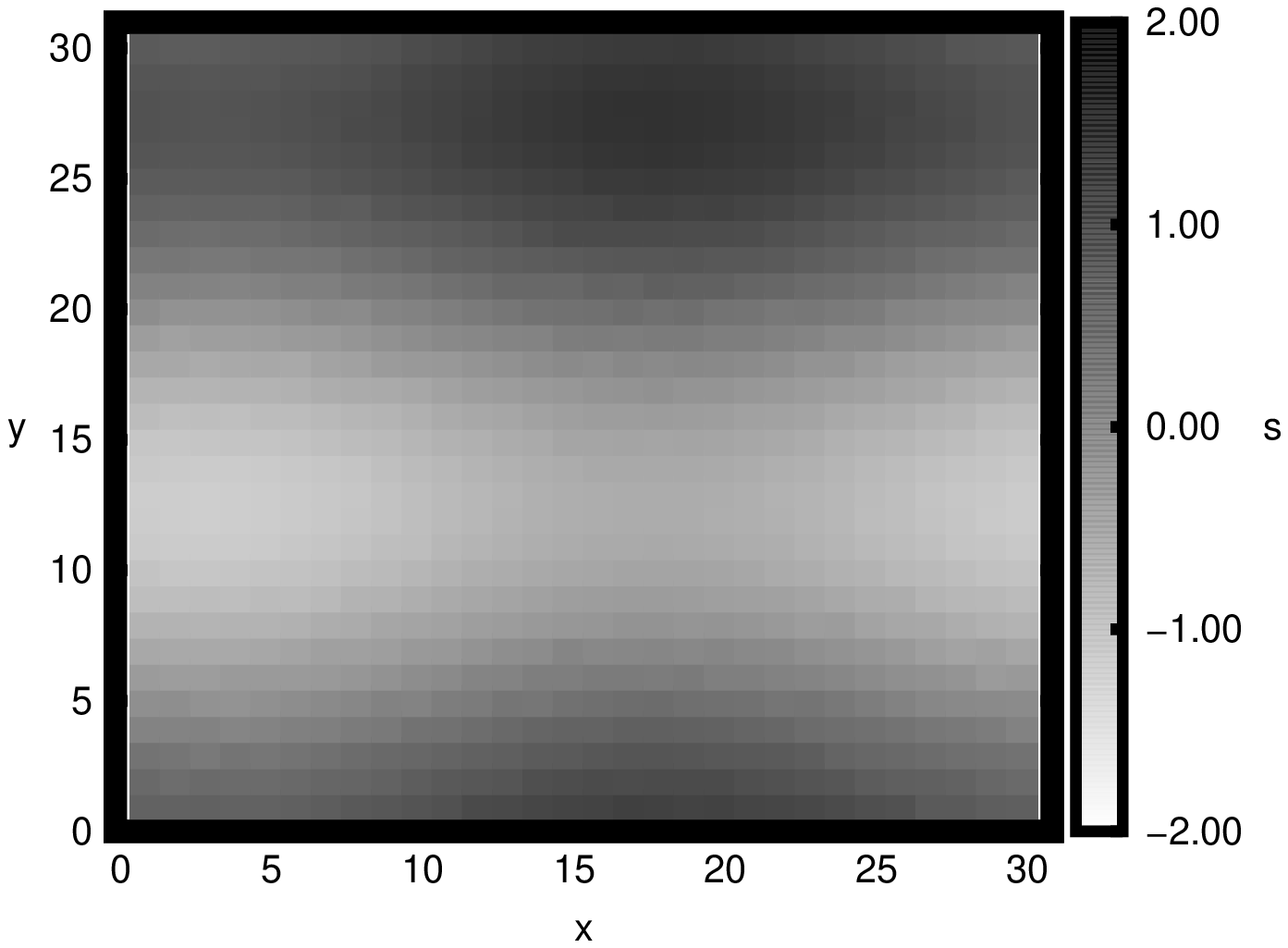}
  \caption{All plots as in Fig.~\ref{fig:6} but with \(\alpha=-0.9\).}
  \label{fig:7}
\end{figure}

In the SnH\(_\text{O}\) region, the system reaches a spatially non-homogeneous state with the agents divided into two groups of different well-defined opinions, and most of the cases, with circular symmetry. This can be seen in the bottom-right plots of Figs.~\ref{fig:6} (for \(\alpha=0.96\)) and Figs.~\ref{fig:7} (for \(\alpha=-0.9\)), both with \(\beta=0\). For these cases (\(N=900\)), the critical value of \(\alpha\), given by Eq.~\eqref{eq:crital}, is \(\alpha_c\simeq 0.978\). The evolution of both kurtosis values (bottom-left plots of Figs.~\ref{fig:6} and \ref{fig:7}) is similar: it increases up to \(10^2t_0\) and then decreases towards a plateau at around \(10^3t_0\). The first evolution stage is interpreted as a transient involving local changes mostly, while the decreasing evolution is associated to the formation of non-homogeneities. In the last stage, the clusters of agents with similar opinions are destroyed, created, and move.   

Finally, in SnH\(_\text{L}\) the agents tend to distribute spatially non-homogeneously, but also to form filaments locally, with L-structures (see the top right graphs of Figs.~\ref{fig:5} and \ref{fig:6}). Now, the temporal evolution of the corresponding kumulants is different. For \(\alpha=0.96\) (relatively close to \(\alpha_c\)), the kumulant first decreases and then increases until reaching a final plateau. In the first stage, filaments form, and subsequently they move and organize non-homogeneously. For \(\alpha=-0.9\), Fig.~\ref{fig:6}, the formation of inhomogeneities occurs earlier and novel structures emerge. This makes the dynamics very slow and makes it difficult to observe a final relaxation to a configuration similar to that in the top-right graph of Fig.~\ref{fig:5}.

\section{Discussion and conclusions\label{sec:diss}}

In this work, we extended the study of a continuous opinion model proposed in Ref.~\cite{khalil2021approach} by considering more complex social interactions. While in all of them the system reaches consensus, the way of approaching it depends critically on both the system parameters and the topology of the interactions. We consider these results particularly relevant for understanding the dynamics of other social systems driven by weak and/or infrequent social encounters; that is, when the outcome of a social interaction produces small changes in the opinions and/or the probability of an interaction is small, such that relaxation towards an eventual final state is difficult to achieve and only part of the relaxation is observable.

Three different ways of approaching consensus have been identified. When agents are well mixed, not only with the all-to-all configuration (for any values of the dynamic parameters), but also within regular lattices (for restricted values of the parameters), the system approaches consensus homogeneously, meaning that all agents keep having the same opinion distribution for all times, hence the same mean opinion and opinion temperature (or opinion dispersion). Interestingly, the opinion distribution obeys a scaling law in which all time dependence occurs through the opinion temperature, similar to the scaling law of the homogeneous cooling state (HCS) of a granular gas \cite{brey1996homogeneous,jimenez2024fluctuations}. Moreover, the scaled opinion distribution can either be unimodal or bimodal, depending on the system parameters, extending the findings of Refs.~\cite{khalil2021approach,carro2013role,khalil2018generalized} to other social conditions. 

In a second scenario, agents may approach to consensus (weak) non-homogeneously. A weak non-homogeneous approach has been observed when the interaction networks are of Erdös-Rényi and Barabási-Albert: all agents have the same zero mean opinion (the initial one) but the opinion dispersion (temperature) can differ. This means that all agents have the same opinion on average, but some ones are more prone to change them that others, depending on the local structure of their interactions. It has been shown that this behaviour is compatible with the exact dynamics identified in Sec.~\ref{sec:lattice}, provided the temperatures depend on time through the same factor. This has been confirmed numerically by finding that agents with different degree have different temperatures, and even different shapes (unimodal versus bimodal) opinion distribution. Interestingly, agents with more connections (neighbours) tend to have less opinion dispersion (temperature), and hence a more robust state, than agents with fewer neighbours.

The approach to consensus can be seen as an equilibration process. In the homogeneous case, it is characterized by all agents having the same mean opinion and the same temperature. In the weak non-homogeneous approach, all opinions have the same mean but now what is common to all agents is the cooling rate (or rate of the change of the temperatures) rather than a common temperature. This modification of the equilibration conditions, compared to the classical ones (same temperatures), is due to the non-equilibrium nature of the dynamics and has been observed in granular gases as well \cite{brey2010critical,brey2011equilibration}. Here, the interaction network seems to play an important role to the new equilibration to be possible. An interesting open question is whether similar equilibration processes can be observed in experimental setups of disordered dense grains where interactions takes place heterogeneously \cite{lechenault2010equilibration,baule2018edwards}.

Still another way of approaching consensus has been observed numerically. If agents interact in a square 2D regular lattice, the spatially homogeneous configuration become unstable when the coefficient \(\alpha\), that modulates the strength of the social interactions, is smaller than a critical value (that depends on the system size). Clusters of agents of similar mean opinions form and move while the overall dispersion of opinions decrease (i.e. while approaching consensus). The role played by the other parameters \(\beta\) (tuning the interaction frequency) is mainly to modify the opinion local structures of the interacting agents. The parameter space \((\alpha,\beta)\) is split into different regions in which the system exhibits different geometrical properties, see Fig.~\ref{fig:4}.

The instability of homogeneous solutions in the lattice configuration presents similarities with that of the homogeneous cooling states of granular gases \cite{brey2008shear,ruiz2009development,fullmer2017clustering}, especially for \(\alpha>0\). Excluding high values of \(\beta\) (not analyzed numerically in this work), in both cases the appearance of spatially non-homogeneous solutions is due to the instability of the velocity mode, and depend on the system size. For \(\alpha\) close to \(-1\), the dynamical evolution of the instabilities in our model becomes very slow and some transient and metastable configurations, such as the one in Fig.~\ref{fig:7}, resemble the pattern formation of agent-based models with long-range interactions \cite{hernandez2004clustering,khalil2017nonlocal}. However, the complete analogy is not yet clear.

Finally, it is worth mentioning that most results of the present work can be extrapolated to another models with additional dynamical ingredients, like radicalization \cite{li2011trust,alizadeh2014distributions,galam2016modeling}. Namely, the main reason the current model described an approach to consensus is that the social compromise propensity is the only included mechanism for the opinion evolution. However, if we take into account that agents may also become intrinsically more radical, by increasing her absolute value of her opinion in absolute value as time rises, then a non-equilibrium steady state can be reached which can be of no consensus, in general. We can guess the results by comparing the present models with similar ones studied in the context of granular gases. Depending on precisely how the two mechanisms are combined, the resulting model can be exactly mapped to the present one \cite{lutsko2001model,brey2004steady}, or give rise to other phenomenology \cite{khalil2020unified}.

\section*{Acknowledgments}

Partial financial support has been received from Grant No. PID2023-151960NA-I00 funded by Ministerio de Ciencia, Innovaci\'on y Universidades (Spain), Agencia Estatal de Investigaci\'on (AEI, Spain, Grant No. 10.13039/501100011033) and by the Community of Madrid and Rey Juan Carlos University through the Young Researchers program in R\&D (Grant No. CCASSE M2737).

\appendix

\section{Details on the derivation of the hydrodynamic description \label{appen:1}}

\subsection{Balance equations\label{appen:11}}

The usual hydrodynamic fields in kinetic theory are the first three moments of \(p(\bs r,s,t)\). In this case, they are the number density of agents \(n(\bs r,t)\) and the already defined local mean opinion \(S(\bs r,t)\) and opinion temperature \(T(\bs r,t)\). They can be directly computed from \(p(\bs r,s,t)\) as
\begin{eqnarray}
  && n(\bs r,t)=\int ds\, p(\bs r,s,t), \\
  && n(\bs r,t)S(\bs r,t)=\int ds\, s p(\bs r,s,t), \\
  && n(\bs r,t)T(\bs r,t)=\int ds\, (s-S)^2 p(\bs r,s,t).
\end{eqnarray}

In the present model, agents do not move and keep homogeneously distributed. Hence, the field \(n(\bs r,t)\) is irrelevant. Namely, from the normalization of \(p(\bs r,s,t)\) it follows that \(n(\bs r,t)=1\). Consistency requires that the kinetic equation \eqref{eq:kineticeq} preserve this constraint. For checking that, the following property of the right-hand side of the kinetic equation \eqref{eq:kineticeq} will be very useful:
\begin{eqnarray}
  \label{eq:prop1}
  && \int ds_1\,f(s_1)\int ds_2\,(|\alpha|^{-1}b_{12}^{-1}-1)\pi(s_1,s_2)p(\bs r,s_1,t)\nabla^2 p(\bs r,s_2,t)\nonumber \\ &&=\iint ds_1ds_2 \pi(s_1,s_2)p(\bs r,s_1,t)\nabla^2 p(\bs r,s_2,t)(b_{12}-1)f(s_1),
\end{eqnarray}
valid for any function \(f(s_1)\) and results after a change of opinion variable. A similar expression is also valid for the ``traditional'' collision term, the first one on the right-hand side of Eq.~\eqref{eq:kineticeq}. Now, integrating Eq.~\eqref{eq:kineticeq} over \(s_1\) and using relation \eqref{eq:prop1} with \(f=1\) we have
\begin{equation}
  \partial_t n(\bs r,t)=0.
\end{equation}
The solution to the previous equation for the initial condition \(n(\bs r,0)=1\) is \(n(\bs r,t)=1\) for all \(\bs r\) and \(t\ge 0\), as desired.

As for the mean opinion, using relation \eqref{eq:prop1} with \(f=s_1\) and that \((b_{12}-1)s_1=\frac{1+\alpha}{2}(s_2-s_1)\), we have
\begin{eqnarray}
\partial_t S(\bs r,t)&=&d(1+\alpha)  \iint ds_1 ds_2 (s_2-s_1)\pi(s_1,s_2)p(\bs r,s_1,t)p(\bs r,s_2,t)\nonumber \\ &&+\frac{1+\alpha}{2} \iint ds_1ds_2(s_2-s_1)\pi(s_1,s_2)p(\bs r,s_1,t)\nabla^2 p(\bs r,s_2,t)\nonumber \\ &=&\frac{1+\alpha}{2} \iint ds_1ds_2(s_2-s_1)\pi(s_1,s_2)p(\bs r,s_1,t)\nabla^2 p(\bs r,s_2,t), 
\end{eqnarray}
where we have used that
\begin{equation}
  \iint ds_1 ds_2 (s_2-s_1)\pi(s_1,s_2)p(\bs r,s_1,t)p(\bs r,s_2,t)=0,
\end{equation}
since \(\pi(s_1,s_2)=\pi(s_2,s_1)\) and, after the change of variables \(\{s_1,s_2\}\to\{s_2,s_2\}\), the previous integral changes its sign. Using now that 
\begin{equation}
  p(\bs r,s_1,t)\nabla^2 p(\bs r,s_2,t)=\nabla\cdot[p(\bs r,s_1,t)\nabla p(\bs r,s_2,t)]-\nabla p(\bs r,s_1,t)\cdot \nabla p(\bs r,s_2,t)
\end{equation}
and
\begin{eqnarray}
  && \iint ds_1ds_2(s_2-s_1)\pi(s_1,s_2) \nabla p(\bs r,s_1,t)\cdot \nabla p(\bs r,s_2,t) \nonumber \\ &&=-\iint ds_1ds_2(s_2-s_1)\pi(s_1,s_2) \nabla p(\bs r,s_1,t)\cdot \nabla p(\bs r,s_2,t)=0,
\end{eqnarray}
we have
\begin{eqnarray}
\partial_t S(\bs r,t)&=&\frac{1+\alpha}{2} \nabla \cdot \iint ds_1ds_2(s_2-s_1)\pi(s_1,s_2)p(\bs r,s_1,t)\nabla p(\bs r,s_2,t). 
\end{eqnarray}
This is not a diffusion equation except for very specific forms of \(\pi(s_1,s_2)\), see for instance the case \(\beta=0\) below.

Proceeding similarly with the opinion temperature, we have:
\begin{eqnarray}
  \partial_tT(\bs r,t)+2S(\bs r,t)\partial_tS(\bs r,t)=-\frac{d(1-\alpha^2)}{2}\iint ds_1ds_2(s_2-s_1)^2\pi(s_1,s_2)p(\bs r,s_1,t) p(\bs r,s_2,t) && \nonumber \\ \qquad +(1+\alpha)\iint ds_1ds_2(s_2-s_1)\pi(s_1,s_2)\left[s_1+\frac{1+\alpha}{4}(s_2-s_1)\right]p(\bs r,s_1,t)\nabla^2 p(\bs r,s_2,t). &&
\end{eqnarray}

\subsection{Case \(\beta =0\)\label{appen:12}}

For \(\beta=0\) it is \(\pi=1\) and the integrals of the balance equations for \(S\) and \(T\) can be explicitly done. After some computation, we have
\begin{eqnarray}
  \label{eq:eqSMax}
  && \partial_t S(\bs r,t)=\frac{1+\alpha}{2}\nabla^2S(\bs r,t),\\
  \label{eq:eqTMax}
  && \partial_t T(\bs r,t)=-d(1-\alpha^2)T+\frac{(1+\alpha)^2}{4}\nabla^2T(\bs r,t)+\frac{(1+\alpha)^2}{2}[\nabla S(\bs r,t)]^2.
\end{eqnarray}
Notably, the set of equations is closed and can be solved given an initial and boundary conditions.

The case \(\beta>0\) is more involved and can not be addressed exactly. Approximate closed hydrodynamic equations will be obtain next. 

\subsection{Chapman-Enskog method \label{appen:13}}

In order to obtain closed hydrodynamic equations for any \(\beta \ge 0\), we follow the Chapman-Enskog method. We seek a normal solution \(p(\bs r,s,t)\), which is a special solution to the kinetic equation that depends on space and time through a functional dependence of the hydrodynamic fields \(S(\bs r,t)\) and \(T(\bs r,t)\). Moreover, the normal solution is found in a perturbative way, by assuming 
\begin{eqnarray}
  && p=p_0+\epsilon p_1+\dots, \\
  && \partial_t=\partial_t^{(0)}+\epsilon \partial_t^{(1)}+\dots, \\
  && \nabla \sim \epsilon,
\end{eqnarray}
with \(\epsilon\) a nonuniformity parameter. The hydrodynamic fields are assumed to be of zeroth order in \(\epsilon\). Using the expansions with the kinetic equation, and equating the coefficients of the different powers of \(\epsilon\), we get up to second order:
\begin{eqnarray}
  && \partial_t^{(0)}p_0= 2d \int ds_2\,(|\alpha|^{-1}b_{12}^{-1}-1)\pi(s_1,s_2)p_0(\bs r_i,s_1,t)p_0(\bs r_i,s_2,t),\\
  && \partial_t^{(0)}p_1+\partial_t^{(1)}p_0= 2d \int ds_2\,(|\alpha|^{-1}b_{12}^{-1}-1)\pi(s_1,s_2)[p_1(\bs r_i,s_1,t)p_0(\bs r_i,s_2,t)+p_0(\bs r_i,s_1,t)p_1(\bs r_i,s_2,t)],\\
  && \partial_t^{(0)}p_2+\partial_t^{(1)}p_1+\partial_t^{(2)}p_0= 2d \int ds_2\,(|\alpha|^{-1}b_{12}^{-1}-1)\pi(s_1,s_2)[p_2(\bs r_i,s_1,t)p_0(\bs r_i,s_2,t)+p_1(\bs r_i,s_1,t)p_1(\bs r_i,s_2,t)\nonumber \\
  && \qquad +p_0(\bs r_i,s_1,t)p_2(\bs r_i,s_2,t)] + \int ds_2\,(|\alpha|^{-1}b_{12}^{-1}-1)\pi(s_1,s_2)p_0(\bs r_i,s_1,t)\nabla^2 p_0(\bs r_i,s_2,t).
\end{eqnarray}

The first equation, for \(p_0\), is essentially the local version of the kinetic Boltzman equation of mean field, analysed in \cite{khalil2021approach}. Its solution can be written as
\begin{equation}
  p_0(\bs r,s,t)=s_0^{-1}\phi(c),
\end{equation}
with
\begin{eqnarray}
  && s_0=\sqrt{2T(\bs r,t)}, \\
  && c=\frac{s-S(\bs r,t)}{s_0},
\end{eqnarray}
and \(\phi\) the solution to Eq.~(57) in \cite{khalil2021approach}. As a first approximation we take
\begin{equation}
  \phi(c)\simeq \pi^{-\frac12}e^{-c^2},
\end{equation}
the so-called Gaussian approximation.

Once \(p_0\) is known, we proceed with \(p_1\). From the scaling properties of \(p_0\) it follows 
\begin{equation}
  \partial_t^{(1)}p_0=\partial_Sp_0 \partial_t^{(1)}S+\partial_Tp_0 \partial_t^{(1)}T,
\end{equation}
where
\begin{eqnarray}
  && \partial_S p_0=\partial_S (s_0^{-1}\phi(c))=-s_0^{-2}\phi'(c),\\
  && \partial_T p_0=-s_0^{-3}[c\phi(c)]',
\end{eqnarray}
where the prime \('\) indicates derivative, and \(\partial_t^{(1)}S\) and \(\partial_t^{(1)}T\) can be computed from the balance equations \eqref{eq:balanceS}--\eqref{eq:balanceT} of \(S\) and \(T\) to order \(\epsilon\):
\begin{eqnarray}
  &&\partial_t^{(1)}S=0,\\
  &&\partial_t^{(1)}T=-\frac{d(1-\alpha^2)}{2}\iint ds_1ds_2(s_2-s_1)^2\pi(s_1,s_2)[p_1(\bs r,s_1,t) p_0(\bs r,s_2,t)+p_0(\bs r,s_1,t) p_1(\bs r,s_2,t)].
\end{eqnarray}
Note that the last expression is a functional of \(p_1\). This way
\begin{eqnarray}
  && \partial_t^{(0)}p_1-\frac{d(1-\alpha^2)}{2}\partial_Tp_0\int ds_1ds_2\,(s_2-s_1)^2\pi(s_1,s_2)[p_1(\bs r,s_1,t) p_0(\bs r,s_2,t)+p_0(\bs r,s_1,t) p_1(\bs r,s_2,t)] \nonumber \\ && = 2d \int ds_2\,(|\alpha|^{-1}b_{12}^{-1}-1)\pi(s_1,s_2)[p_1(\bs r_i,s_1,t)p_0(\bs r_i,s_2,t)+p_0(\bs r_i,s_1,t)p_1(\bs r_i,s_2,t)],
\end{eqnarray}
which is ``linear'' in \(p_1\) and has
\begin{equation}
  p_1=0
\end{equation}
as a solution. We assume this solution is the relevant one at the hydrodynamic scales, which is reached very quickly (with respect to the hydrodynamic time scales) from any initial condition. Hence, the first-order contribution to the opinion distribution can be neglected.

Finally, if we want a system of hydrodynamic equations up to order \(\epsilon^2\), the so-called Navier-Stokes order, we should consider the equation for \(p_2\). However, it is easily seen that the contribution of \(p_2\) to the resulting equations is only through a correction to the leading order of the cooling rate, Eq.~\eqref{eq:coolingr} after approximating \(p_{ij}\), see below. By neglecting the contribution of \(p_2\), the Navier-Stokes equations read
\begin{eqnarray}
  \label{eq:ns1}
  && \partial_t S(\bs r,t)\simeq \frac{1+\alpha}{2} \nabla \cdot \iint ds_1ds_2(s_2-s_1)\pi(s_1,s_2)p_0(\bs r,s_1,t)\nabla p_0(\bs r,s_2,t), \\
  \label{eq:ns2}
  && \partial_tT(\bs r,t)+2S(\bs r,t)\partial_t S(\bs r,t)\simeq -\frac{d(1-\alpha^2)}{2}\iint ds_1ds_2(s_2-s_1)^2\pi(s_1,s_2)p_0(\bs r,s_1,t) p_0(\bs r,s_2,t) \nonumber \\ && \qquad +(1+\alpha)\iint ds_1ds_2(s_2-s_1)\left[s_1+\frac{1+\alpha}{4}(s_2-s_1)\right]\pi(s_1,s_2)p_0(\bs r,s_1,t)\nabla^2 p_0(\bs r,s_2,t),
\end{eqnarray}
where \(p_0\) is supposed to be known. Two remarks are in order. On the one hand, the previous set of equations corresponds to a ``local equilibrium'' approximation of the ``exact'' balance equations \eqref{eq:balanceS}--\eqref{eq:balanceT}; that is, the set \eqref{eq:balanceS}--\eqref{eq:balanceT} with the distribution of opinions \(p\) replaced by \(p_0\) (the local version of the distribution of opinions in a homogeneous approach to consensus). On the other hand, the mathematical structure of the hydrodynamic equations is different from the case of granular gases, see for instance \cite{khalil2014hydrodynamic}. Namely, here the contribution to the fluxes are purely collisional, due to the interactions between agents. The important contributions due to the stream motion of grains are absent here: agents ``keep at rest''.

We now compute the integrals in Eqs.~\eqref{eq:ns1}--\eqref{eq:ns2}. Consider first the equation of \(S\). Using the form of \(\pi(s_1,s_2)\), see Eq.~\eqref{eq:pis1s2}, and
\begin{equation}
  \nabla p_0=\partial_Sp_0\nabla S+\partial_Tp_0\nabla T=-s_0^{-2}\phi'\nabla S-s_0^{-3}\left(c\phi\right)'\nabla T
\end{equation}
we have 
\begin{eqnarray}
  \partial_t S(\bs r,t) &\simeq& -\frac{1+\alpha}{2} \nabla \cdot \left[\frac{s_0^\beta}{\Delta s_t^\beta}\iint dc_1dc_2|c_2-c_1|^\beta (c_2-c_1)\phi(c_1)\phi'(c_2)\nabla S(\bs r,t)\right] \nonumber \\ && -\frac{1+\alpha}{2} \nabla \cdot \left[\frac{s_0^{\beta-1}}{\Delta s_t^\beta}\iint dc_1dc_2|c_2-c_1|^\beta (c_2-c_1)\phi(c_1)[c_2\phi(c_2)]'\nabla T(\bs r,t)\right].
\end{eqnarray}
Since \(\phi(c)\) is an even function of \(c\), the second integral in the previous expression vanishes. The equation for \(S\) can then be written as a diffusion equation, Eq.~\eqref{eq:diffS}, with the diffusion coefficient \(D\) given by 
\begin{eqnarray}
  D(\bs r,t)&=&\frac{1+\alpha}{4}\frac{[2T(\bs r,t)]^{\frac{\beta}{2}}}{\Delta s_t^\beta} \iint dc_1dc_2 (c_2-c_1)|c_2-c_1|^\beta [\phi'(c_1)\phi(c_2)-\phi(c_1)\phi'(c_2)] \nonumber \\ &\simeq & \frac{2^{\beta}(1+\alpha)}{\sqrt{\pi}}\Gamma\left(\frac{\beta+3}{2}\right) \frac{[T(\bs r,t)]^{\frac{\beta}{2}}}{\Delta s_t^\beta},
\end{eqnarray}
where the last equality uses the Gaussian approximation for \(\phi\). We recall that \(\Delta s_t\), which may also depend on time, is a given quantity of the model.

As for the equation for \(T\), the first integral in Eq.~\eqref{eq:ns2} is related to the cooling rate \(\zeta\):   
\begin{eqnarray}
  \label{eq:zeta}
  \zeta(\bs r,T)T(\bs r,t)&\simeq & \frac{d(1-\alpha^2)}{2}\iint ds_1ds_2(s_2-s_1)^2\pi(s_1,s_2)p_0(\bs r,s_1,t) p_0(\bs r,s_2,t) \nonumber \\ &=& \frac{2^{\beta+1}d(1-\alpha^2)}{\sqrt{\pi}}\frac{s_0^{\beta+2}}{\Delta s_t^\beta}\iint dc_1dc_2|c_2-c_1|^{\beta+2}\phi(c_1)\phi(c_2) \nonumber \\ & \simeq & \frac{2^{\beta+1}d(1-\alpha^2)}{\sqrt{\pi}}\Gamma\left(\frac{\beta+3}{2}\right)\frac{[T(\bs r,t)]^{\frac{\beta}{2}+1}}{\Delta s_t^\beta},
\end{eqnarray}
where the last approximate equality holds under the Gaussian approximation for \(\phi\). Using that \(p_0(\bs r,s_1,t)\nabla^2 p_0(\bs r,s_2,t)=\nabla\cdot [p_0(\bs r,s_1,t)\nabla p_0(\bs r,s_2,t)]-\nabla p_0(\bs r,s_1,t)\cdot \nabla p_0(\bs r,s_2,t)\), the other integral in the equation for \(T\) can be written as
\begin{eqnarray}
  && (1+\alpha) \nabla \cdot \int ds_1ds_2\,(s_2-s_1)\left[s_1+\frac{1+\alpha}{4}(s_2-s_1)\right]\pi(s_1,s_2)p_0(\bs r,s_1,t)\nabla p_0(\bs r,s_2,t) \nonumber \\ && -(1+\alpha) \int ds_1ds_2\,(s_2-s_1)\left[s_1+\frac{1+\alpha}{4}(s_2-s_1)\right]\pi(s_1,s_2)\nabla p_0(\bs r, s_1,t)\cdot\nabla p_0(\bs r,s_2,t) \\
  && =-\frac{1+\alpha}{\Delta s_t^\beta}\nabla\cdot\left[s_0^{\beta+1}\left(I_S\nabla S+\frac{I_T}{s_0}\nabla T\right)\right]+\frac{1-\alpha^2}{4\Delta t^\beta}s_0^\beta \left[I_{SS}(\nabla S)^2+\frac{I_{ST}}{S_0}\nabla S\cdot \nabla T+\frac{I_{TT}}{s_0^2}(\nabla T)^2\right],
\end{eqnarray}
where use has been made of the scaling form of \(p_0\). The new quantities are the following integrals:
\begin{eqnarray}
  I_S&=&\iint dc_1dc_2|c_1-c_2|^\beta \left[c_1+\frac{S(\bs r,t)}{s_0}+\frac{1+\alpha}{4}(c_2-c_1)\right]\phi(c_1)\phi'(c_2)\simeq -\frac{2^{\frac{\beta}{2}+1}}{\sqrt{\pi}}\Gamma\left(\frac{\beta+3}{2}\right)\frac{S(\bs r,t)}{s_0(\bs r,t)}, \\
  I_T&=&\iint dc_1dc_2|c_1-c_2|^\beta \left[c_1+\frac{S(\bs r,t)}{s_0}+\frac{1+\alpha}{4}(c_2-c_1)\right]\phi(c_1)[c_2\phi(c_2)]'\nonumber \\ &\simeq& -\frac{2^{\frac{\beta}{2}}}{\sqrt{\pi}}\Gamma\left(\frac{\beta+3}{2}\right)\left[\frac{(1+\alpha)(2+\beta)}{4}-\frac{\beta}{2}\right], \\
  I_{SS}&=&\iint dc_1dc_2|c_2-c_1|^{\beta+2}\phi'(c_1)\phi'(c_2)\simeq -\frac{2^{\frac{\beta}{2}+1}(2+\beta)}{\sqrt{\pi}}\Gamma\left(\frac{\beta+3}{2}\right), \\
  I_{ST}&=&\iint dc_1dc_2|c_2-c_1|^{\beta+2}\left\{\phi'(c_1)[c_2\phi(c_2)]'+[c_1\phi(c_1)]'\phi'(c_2)\right\}=0, \\
    I_{TT}&=&\iint dc_1dc_2|c_2-c_1|^{\beta+2}[c_1\phi(c_1)]'[c_2\phi(c_2)]\simeq -\frac{2^{\frac{\beta}{2}-1}\beta (2+\beta)}{\sqrt{\pi}}\Gamma\left(\frac{\beta+3}{2}\right),
\end{eqnarray}
evaluated, again, using the Gaussian approximations. Rearranging the different terms and using the equation for \(S(\bs r,t)\), the resulting equation for the temperature can be written as Eq.~\eqref{eq:eqtemp} with the new transport coefficients given by
\begin{eqnarray}
  \label{eq:kappa}
  && \kappa(\bs r,t)\simeq \frac{2^\beta(1+\alpha)}{\sqrt{\pi}}\Gamma\left(\frac{\beta+3}{2}\right)\left[\frac{(1+\alpha)(2+\beta)}{4}-\frac{\beta}{2}\right]\frac{T^{\frac{\beta}{2}}}{\Delta s_t^\beta}, \\
  \label{eq:etaS}
  && \eta_S(\bs r,t)\simeq \frac{2^\beta(1+\alpha)}{\sqrt{\pi}}\Gamma\left(\frac{\beta+3}{2}\right)\left[2-\frac{(1-\alpha)(2+\beta)}{2}\right]\frac{T^{\frac{\beta}{2}}}{\Delta s_t^\beta}, \\
  \label{eq:etaT}
  && \eta_T(\bs r,t)\simeq \frac{2^{\beta-4}(1-\alpha^2)}{\sqrt{\pi}}\beta(2+\beta)\Gamma\left(\frac{\beta+3}{2}\right)\frac{T^{\frac{\beta}{2}-1}}{\Delta s_t^\beta},
\end{eqnarray}
withing the Gaussian approximation. Note that not all transport coefficient have a definite sign for all values of the parameters \(\alpha\) and \(\beta\). In particular, \(\kappa\), which can be thought as a thermal conductivity, changes sign its for \(\alpha\simeq \frac{\beta-2}{\beta+2}\).

An important observation is that the Navier-Stokes equations that result from the Chapman-Enskog method up to second order in the gradients, using the Gaussian approximation, reproduce the ``exact'' result for \(\beta=0\), Eqs.~\eqref{eq:eqSMax}--\eqref{eq:eqTMax}. 
 
\bibliographystyle{elsarticle-num} 
\bibliography{vgr_bib}

\end{document}